\documentclass[superscriptaddress,aps,pra,twocolumn,preprintnumbers,amsmath,amssymb,floatfix,showpacs,10pt]{revtex4-2}
\usepackage[utf8x]{inputenc}

\usepackage{graphics}
\usepackage{epsfig}
\usepackage{subfigure}
\usepackage{rotating}
\usepackage{mathrsfs}
\usepackage{amsfonts}
\usepackage{amsmath}
\usepackage{amssymb}
\usepackage{placeins}
\usepackage{bm}

\usepackage{pgfplots}
\usepackage{tikz}
\usetikzlibrary{arrows,decorations.markings}
\pgfplotsset{compat=1.14}

\usepackage{soul}
\usepackage{fixmath}
\usepackage{physics}

\usepackage[colorlinks]{hyperref}

\begin{document}

\title{A theoretical framework for {photon-subtraction with non-mode selective resources}}
\author{M. F. Melalkia}
\author{L. Brunel}
\author{S. Tanzilli}
\author{J. Etesse}

\affiliation{
Universit\'{e} Cote d’Azur, CNRS, Institut de Physique de Nice, Parc Valrose, 06108 Nice Cedex 2, France
}

\author{V. D'Auria}

\affiliation{
Universit\'{e} Cote d’Azur, CNRS, Institut de Physique de Nice, Parc Valrose, 06108 Nice Cedex 2, France
}

\affiliation{Institut Universitaire de France (IUF), France}

\begin{abstract}
This work establishes a versatile theoretical framework that explicitly describes single-photon subtraction from multimode quantum light in the context of non-Gaussian state generation and manipulation. The treatment focuses on easy-to-implement configurations in which no mode-selective operation is available and evaluates features and advantages of scheme where only simple filtering stages are employed on the experiments. Such configuration, by considerably reducing the experimental overheads, makes experiments involving single photon subtraction easier to be implemented. Obtained theoretical framework allows retrieving, given a multimode input state, optimal conditions required to herald and then to detect non-Gaussian states, providing a practical and powerful toolbox for experiments' design. The application of the proposed approach to the case study of Schr\"odinger kitten preparation starting from a frequency multimode squeezed state illustrates the impact of the derived theoretical tools.
\end{abstract}

\date{\today}
\maketitle
\section*{Introduction}
In the context of emerging quantum technologies, continuous variable (CV) quantum optics offers a wide panel of applications, encompassing quantum metrology~\cite{Giovannetti2006}, quantum communication~\cite{Braunstein2005a}, and quantum computing~\cite{Ferraro_Book}. A full exploitation of the CV approach relies on the ability of producing high quality states represented by non-Gaussian functions in the quantum phase space~\cite{ShapiroGaussianQI2012}. A 
powerful strategy to generate and manipulate such non-Gaussian states consists in subtracting (or adding) photons from an appropriate input. In particular, the subtraction of single photons from squeezed vacuum is the most common way to produce key CV resources such as Schr\"odinger kitten- and Fock-states~\cite{dakna1997generating, LvovskyNG2020}, experimentally demonstrated in both pulsed~\cite{ourjoumtsev2006} and continuous-wave~\cite{neergaard2006, wakui2007} regime. At the same time, the performances of this kind of operations are strongly affected by the properties of the protocol input state whose multimode features must be carefully mastered. In experiments, this point can play a very critical role, as most of bulk and guided-wave realisations rely on input squeezed states produced by spontaneous parametric down conversion (SPDC), that, depending on the working conditions, can be highly multimode in the frequency domain~\cite{wasilewski2006pulsed, sasaki2006multimode}. To comply with this situation, single-mode engineered SPDC sources~\cite{Silberhorn2011SingleModeTWB} as well as mode-selective state manipulation~\cite{walschaers2018tailoring, TrepsNG2020} have been studied and experimentally demonstrated. Nevertheless, the specific experimental conditions required for such operations remain difficult or unaccessible to many practical situations. 
\bigskip 

In this context, this work offers a general and versatile theoretical framework able to describe single photon subtraction in a highly multimode context and with no hypothesis on the nature, shape or number of involved modes. Previous theoretical models have already treated the case of mode-selective photon-subtraction schemes~\cite{averchenko2016multimode, walschaers2018tailoring}, whose practical implementation demands non-linear optical stages~\cite{TrepsNG2020,QuantumPulseGate}. This work focuses, instead, on the very common experimental situation in which single-mode or mode-selective operations are not possible or practically unavailable. It aims at assessing the performances of extremely easy-to-implement experimental setups where single-photon subtraction is obtained by using a standard beam-splitter followed by a simple bandpass filter in the heralding path~\cite{LvovskyNG2020}. Thanks to its simplicity, this configurations has been repeatedly implemented in experiments~\cite{ourjoumtsev2006, neergaard2006, wakui2007} but, so far, without an explicit and exhaustive theoretical evaluation of its features. The model presented here closes this gap. Given a multimode state, it makes it possible to derive the exact shape of the state obtained after the single photon subtraction and to determine, based on chosen working parameters, the optimal conditions to announce and detect its non-Gaussian properties. These ingredients are essential for the conception, design and optimisation of any experiments in which multimode features, in frequency, time or space, are important or unavoidable and open the way to future ambitious realisations. As for an illustration of its impact, the developed theoretical framework is applied to Schr\"odinger kitten preparation from a frequency multimode squeezed state, including in the very practical scenario where squeezing is provided by SPDC from a lithium niobate optical waveguide compatible with integrated systems~\cite{Mondain2019}. The properties and explicit shapes of the protocol detection modes are reported and discussed as functions of the input multimode characteristics and of realistic heralding conditions. The developed methods, derived for the generic case of no-mode selective single photon subtraction, can also be applied to further manipulation of non-Gaussian states via cascaded photon subtraction. In addition, as briefly discussed in what follows, they can be easily adapted to the special case of mode-selective operations. Note that theoretical models for mode-selective operations have already been reported in the literature~\cite{averchenko2016multimode, QuantumPulseGate} and, accordingly, shall not be discussed in detail. 
\bigskip

The paper is structured as follows. Section \ref{Photsub} provides a multimode description of single-photon subtraction as well as the explicit expression of the heralded non-Gaussian state as it is measured by a homodyne detection. In Sec. \ref{secCatGen}, the obtained formalism is applied to the case of Schr\"odinger kitten generation. Different experimental configurations for both heralding and detection stages are discussed and compared. For the sake of simplicity, derived results adopt the formalism of multimode features of multiple spectral components. This choice makes it possible to gain easy physical intuitions of obtained results and discuss the common cases of non-Gaussian state generation via single-photon subtraction from squeezed states generated by SPDC~\cite{sasaki2006multimode, averchenko2016multimode}. All numerical simulations refer to the manipulation of squeezing emitted in the C-band of classical telecommunication, compatible with future practical applications to fibre-based quantum communication.

\section{Photon-subtraction on multimode states}
\label{Photsub}
Following a very common strategy, multimode states are treated in terms of the so-called supermodes~\cite{Patera2010}. These permit rewriting the output of a multimode quantum optical source as a tensor product of independent single mode states, each being described by a spectral envelop $\psi_k(\omega)$~\cite{wasilewski2006pulsed, Patera2010, Patera2020}. Supermode envelops $\{\psi_k(\omega)\}$ form an orthonormal basis ($\int \psi_k(\omega)\psi^*_l(\omega)d\omega=\delta_{k,l}$, $\delta_{k,l}$ being the Kronecker delta) and their associated bosonic operators $\{\hat{A_k}\}$  are:
 \begin{equation}\label{Asupermodes}
 \hat{A}_k=\int \psi^*_k(\omega)\, \hat{a}(\omega) \,\mathrm{d}\omega,
 \end{equation}
where $\hat{a}(\omega)$ are the bosonic operators associated with the individual spectral components of the source output.

The multimode state, generically indicated as $\ket{\psi}$, undergoes the typical subtraction scheme shown in Fig.~1; it is sent towards a subtraction beam-splitter (BS) with reflection coefficient $r_s\ll 1$. The reflected beam goes towards the heralding path, while the transmitted one carries the heralded state. In experiments, light in the heralding path is sent to a bucket single-photon detector (SPD), unable to distinguish light from different modes and whose detection signal heralds a successful photon subtraction and the preparation of a desired non-Gaussian state~\cite{LvovskyNG2020}.  At the same time, the subtraction BS, a priori, acts in a similar way on all spectral components of $\ket{\psi}$ or, equivalently, on all supermodes $\{\psi_k(\omega)\}$. This makes it impossible to associate a photon detection event with a subtraction operation on a specific supermode, and can eventually lead to mixed heralded states~\cite{branczyk2010optimized}. To comply with such a situation, a certain mode selectivity on the heralding path is generally obtained by adding an optical frequency filter before the photon-counting detector~\cite{LvovskyNG2020, branczyk2010optimized, christ2014theory, Furusawa2017CWCats}. The filter action can be modelled as a BS: different scenarios can thus be considered based on the shape of the transmission coefficient of the filter BS. \bigskip
\begin{figure} 
    \begin{center}
     \begin{subfigure}
      {\includegraphics[width=\columnwidth]{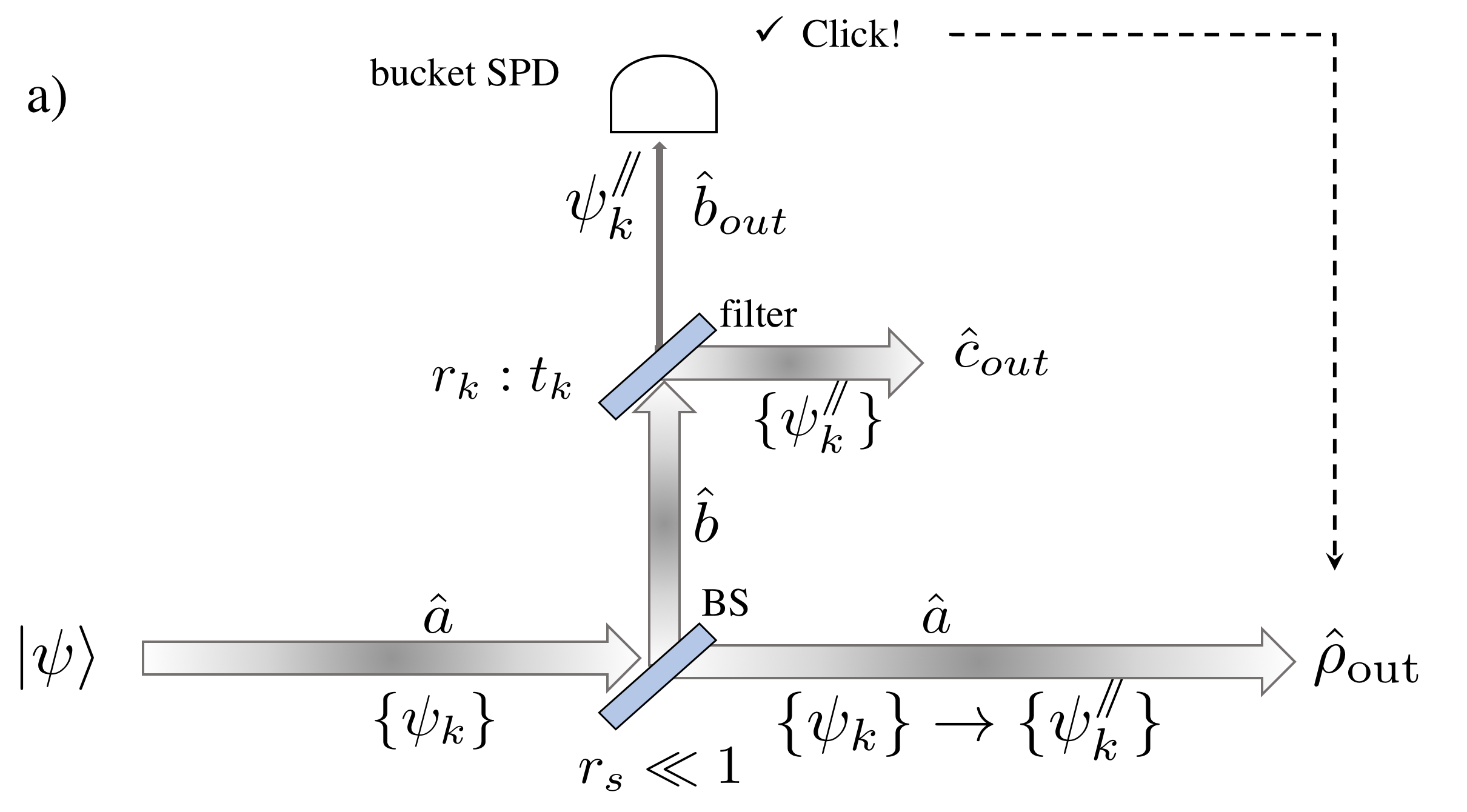}}
     \end{subfigure}
     \vspace{2.5cm}
 \begin{subfigure}
        { \includegraphics[width=\columnwidth]{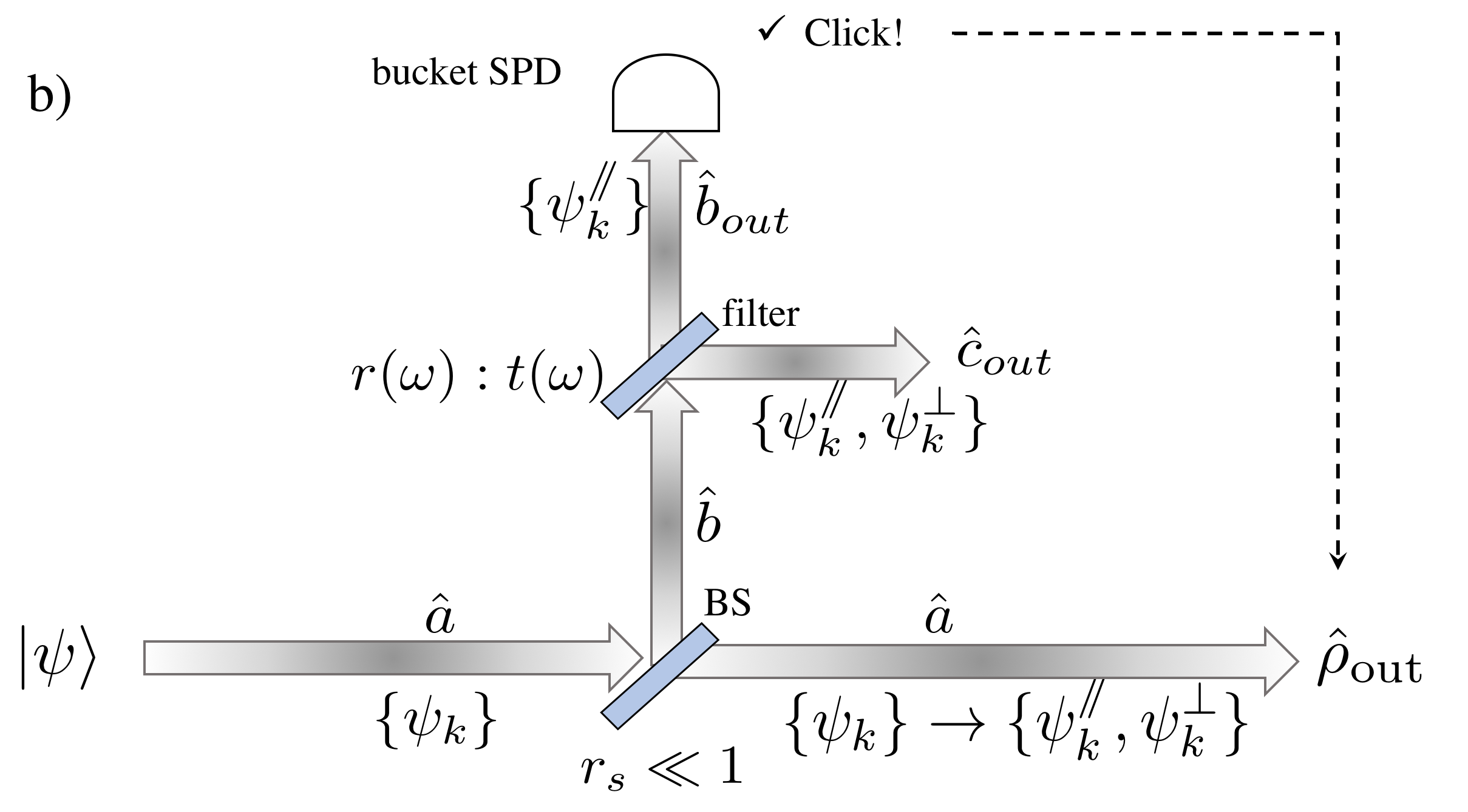}}
     \end{subfigure}
      \end{center}
      \vspace{-3cm}
 \caption{Subtraction scheme applied to a multimode input state $\ket{\psi}$ in case of (a) mode-selective and (b) non mode-selective operations. The working supermode basis for each spatial mode is explicitly indicated in both cases. A hybrid Schr\"odinger / Heisenberg approach is adopted, as the operators do not evolve on the subtracting beam-splitter, but do evolve on the filter beam-splitter. Mode $\hat{b}_{out}$ represents the spatial mode of the heralding photons downstream of the filter that are directed toward the single-photon detector (SPD). Mode $\hat{c}_{out}$ represents photons that are rejected by the filter and thus disregarded. After the photon-counting operation, mode $\hat{a} $ carries the heralded non-Gaussian state  $\hat{\rho}_{\rm out}$. }
\label{subtraction}
\end{figure}

\paragraph*{Perfect mode-selective single-photon subtraction\bigskip\\}
Fig.~\ref{subtraction}-a represents a conceptual scheme of mode-selective photon-subtraction. In this case, the filter is modelled as a mode-selective beam splitter, able to transmit towards the bucket single-photon detector only one given detection mode $\psi_{k}^{/\!\!/}(\omega)$. Depending on the chosen mode, it is possible to manipulate the original multimode state in a controlled manner or even to entangle formerly independent supermodes $\{\psi_{k}(\omega)\}$~\cite{TrepsNG2020}. In particular, in the special case in which the detection mode matches one of the supermode envelops (i.e. $\psi_{k}^{/\!\!/}=\psi_{k})$, Schr\"odinger kitten or Fock states can be heralded. As already discussed, the theory underlying mode-selective operation has already made the object different works~\cite{averchenko2016multimode, QuantumPulseGate} and its treatment is beyond the scope of this paper. At the same time, it is pertinent to observe that original methods and results reported in the following paragraphs can be adapted to the mode-selective case by considering for a filter BS with discrete transmission coefficient $t_k$ equal to 0 for all modes excepted for a specific $\psi_k^{/\!\!/}(\omega)$. \bigskip
\paragraph*{Non mode-selective single-photon subtraction\bigskip\\}
Experimentally, mode-selective single-photon subtractions require adapted non linear optical stages~\cite{TrepsNG2020}. In the large majority of reported experimental works, standard (passive) optical filters are used in the heralding path~\cite{LvovskyNG2020}. In such a non-mode selective operation, the filter can be modelled as a beam-splitter whose real transmission coefficient $t(\omega)$ depends on the optical frequency. No \emph{a priori} hypothesis on the filter transmission profile is imposed.

In order to perform the mathematical derivation, a hybrid Schr\"odinger/Heisenberg approach is adopted~\cite{GattiOriginali2007}: the bosonic operators do not evolve on the subtracting beam-splitter, but do evolve on the filter beam-splitter in the heralding path (see Fig. \ref{subtraction}-a).
Such a strategy allows introducing a global evolution operator, $\hat{\mathcal{U}}$, that can be used to describe the state obtained when both the subtraction and filter beam-splitter are explicitly considered: 
\begin{align}
\hat{\mathcal{U}}=&e^{\theta\int\left(\hat{a}^\dagger(\omega)\hat{b}(\omega)-\hat{a}(\omega)\hat{b}^\dagger(\omega)\right)d\omega}.\\
\approx &\,\hat{1} + \theta\int\left(\hat{a}^\dagger(\omega)\hat{b}(\omega)-\hat{a}(\omega)\hat{b}^\dagger(\omega)\right)d\omega,
\label{BS_sub}
\end{align}
where $\sin \theta/2=r_s \ll 1$ gives the reflection coefficient of the subtraction BS. Spatial modes are named as indicated in Fig.~\ref{subtraction}. The operator $\hat{\mathcal{U}}$ correctly encompasses all the spectral components of the field. 
In the previous expression, each operator $\hat{b}(\omega)$ undergoes the transformation due to the filtering stage and can be rewritten in terms of the bosonic operators $\hat{b}_{out}(\omega)$ and $\hat{c}_{out}(\omega)$ that describe the fields transmitted and reflected by the filter BS, respectively:
\begin{align}\label{boutcout}
\hat{b}(\omega)=t(\omega)\hat{b}_{\rm out}(\omega)+r(\omega)\hat{c}_{\rm out}(\omega),
\end{align}
with $t(\omega)^2+r(\omega)^2=1$. Expression~\eqref{BS_sub} can thus be written as: 
\begin{align}
\hat{\mathcal{U}}\approx\hat{1}+\theta\int &d\omega\left(\hat{a}^\dagger(\omega)t(\omega)\hat{b}_{\rm out}(\omega)+\hat{a}^\dagger(\omega)r(\omega)\hat{c}_{\rm out}(\omega)\right. \nonumber\\
&-\left.\hat{a}(\omega)t^*(\omega)\hat{b}^{\dagger}_{\rm out}(\omega)-\hat{a}(\omega)r^*(\omega)\hat{c}^{\dagger}_{\rm out}(\omega)\right).
\label{Beamsplitt}
\end{align}
To highlight the multimode features, $\hat{\mathcal{U}}$ can be conveniently expressed in terms of bosonic operators associated with the supermodes $\{\psi_k(\omega)\}$ by inverting relation~\eqref{Asupermodes} and by defining, by analogy with $\hat{A}_k$, operators $\hat{B}_{\rm out,k}$ and $\hat{C}_{\rm out,k}$ associated with the spatial modes downstream of the filter BS:
\begin{subequations}\label{decomposition}
\begin{align}
\hat{a}(\omega)&=\sum_k\psi_k(\omega)\hat{A}_k\\
\hat{b}_{\rm out}(\omega)&=\sum_k\psi_k(\omega){\hat{B}_{\rm out,k}}\\
\hat{c}_{\rm out}(\omega)&=\sum_k\psi_k(\omega){\hat{C}_{\rm out,k}}.
\end{align}
\end{subequations}
Transformation~\eqref{decomposition} leads to the appearance in Eq.~\eqref{Beamsplitt} of terms of the kind:
\begin{align}
\hat{a}^\dagger(\omega)t(\omega)\hat{b}_{\rm out}(\omega)=\sum_{k,k'}\psi^*_k(\omega)\hat{A}_k^\dagger t(\omega)\psi_{k'}(\omega)\hat{B}_{\rm out,k'}.
\end{align}
and analogous. However, filtered supermode functions $\{t(\omega)\psi_{k}(\omega)\}$ are no longer appropriate to define optical modes as they are not orthogonal to each other or to the original $\{\psi_{k}(\omega)\}$~\cite{branczyk2010optimized}. This reflects the fact that the contributions of the original input supermodes are mixed together by the mode-insensitive filter. It is thus \emph{a priori} impossible to factorise the action filter BS into that of multiple BS acting each on only a given supermode $\psi_{k}(\omega)$.
\subsection{Supermode decomposition}
\begin{figure*}[t]
\includegraphics[width=2\columnwidth]{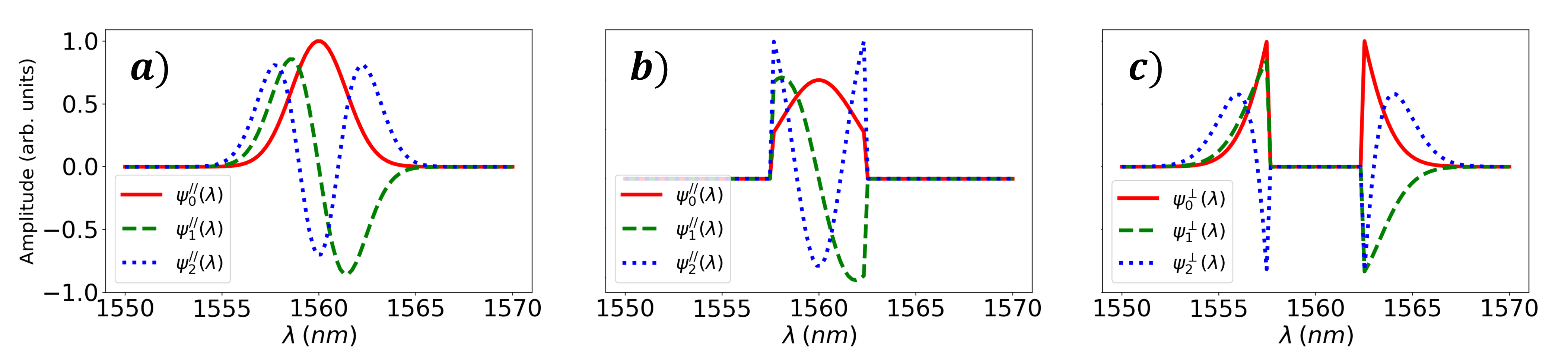}
\caption{(Color online) (a): Examples of the first three $\{\psi_{n}^{/\!\!/}(\omega)\}$ in the case of a filter with a Gaussian transmission coefficient $t(\omega)$ centred around 1560\,nm and with FWHM of 5\,nm ($\approx$\,600\,GHz).  (b)  Examples of the first three $\{\psi_{n}^{/\!\!/}(\omega)\}$ and (c) $\{\psi_{n}^{\perp}(\omega)\},$ in the case of a filter with a rectangular profile centred around 1560\,nm with a FWHM of 5\,nm. For the rectangular filter, the coefficient $t(\omega)$ goes exactly to zero outside the 5\,nm transmission bandwidth. The input state supermodes are assumed to be Hermit functions. For both filters, $\psi_{0}^{/\!\!/}(\omega)$ is given by $\frac{t(\omega)\psi_0(\omega)}{\sqrt{\int \vert t(\omega)\psi_0(\omega)\vert^2 d\omega}}$ with a Gaussian $\psi_0(\omega)$ function. $\{\psi_{n>0}^{/\!\!/}(\omega), \psi_{n}^{\perp}(\omega)\}$ are constructed by Gram-Schmidt process~\cite{nielsen2010quantum}.}\label{ModiParalleliOrtogonali}
\end{figure*}

Products involving the filter transmission and the supermodes functions can adequately be decomposed in a suitable orthonormal basis $\{\psi_{k}^{/\!\!/}(\omega)\}$ whose support corresponds to spectral regions for which $t(\omega)>0$~\cite{branczyk2010optimized}. Thanks to such new modes, it is possible to express the detection of single photons downstream of the filter. However, in principle, the set of $\{\psi_{k}^{/\!\!/}(\omega)\}$ is sufficient \emph{only} to describe the state of systems that actually passed through the filter; as-a-matter-of-fact, in a conditional preparation scheme, this restricts the use of such functions to the sole description of the heralding photon detection. To overcome this limitation, in this work, the basis $\{\psi_{k}^{/\!\!/}(\omega)\}$ is completed with a set of orthonormal functions $\{\psi_{k}^{\perp}(\omega)\}$. Functions $\{\psi_{k}^{/\!\!/}(\omega),\psi_{k}^{\perp}(\omega)\}$ are orthogonal to each other and have disjoint supports. Their properties are detailed in Appendix~\ref{mammt}. The complete orthonormal basis $\{\psi_{k}^{/\!\!/}(\omega),\psi_{k}^{\perp}(\omega)\}$ can now be used to decompose all spectral modes involved in the non-Gaussian manipulation scheme, including those of the subsystem carrying the heralded state. In particular, the supermode functions can be written as:
\begin{align}
\psi_{k}(\omega)&=\sum_n p_{kn}\psi_{n}^{/\!\!/}(\omega)+q_{kn}\psi_{n}^{\perp}(\omega), \label{eq_basis_change}
\end{align}
where
\begin{align}\label{pandp}
p_{kn}&=\int \psi_{k}(\omega)\left(\psi_{n}^{/\!\!/}(\omega)\right)^*d\omega \nonumber \\
q_{kn}&=\int \psi_{k}(\omega)\left(\psi_{n}^{\perp}(\omega)\right)^*d\omega.
\end{align}
The specific shape of  $\{\psi_{k}^{/\!\!/}(\omega),\psi_{k}^{\perp}(\omega)\}$ depends on that of the input supermodes and on the filter profile $t(\omega)$. In the case of continuous functions $t(\omega)$ with continuous derivatives, such as in the case of Lorentzian-shaped or Gaussian filter profiles, as demonstrated in the Appendix~\ref{mammt}, the sets of $\{\psi_{n}^{/\!\!/}(\omega)\}$ can be in principle sufficient to obtain a complete basis (see Eqs.~\eqref{orthmodes_funcinput} and \eqref{psiperp_tran}). In all cases in which $t(\omega)=0$ in one or more spectral regions,  both $\{\psi_{n}^{/\!\!/}(\omega)\}$ and $\{\psi_{n}^{\perp}(\omega)\}$ are non-null and must be taken into account to obtain a complete basis. This concept is illustrated in Fig.~\ref{ModiParalleliOrtogonali} where the first $\{\psi_{n}^{/\!\!/}(\omega)\}$ and $\{\psi_{n}^{\perp}(\omega)\}$ are represented for the case of a Gaussian and of a rectangular filter.

Relations~\eqref{eq_basis_change} can be used to rewrite the bosonic operators associated with the initial system supermodes so as to express single mode squeezed states at the protocol input in terms of the bosonic operators associated with the filter supermodes $\psi_{n}^{/\!\!/}(\omega)$ and $\psi_{n}^{\perp}(\omega)$. More explicitly:
\begin{align}
\hat{A}_k=\sum_np_{kn}^*\hat{\mathcal{A}}_{n}^{/\!\!/}+q_{kn}^*\hat{\mathcal{A}}_{n}^{\perp}\Longleftrightarrow
\begin{array}{c}
\hat{\mathcal{A}}_{n}^{/\!\!/}=\sum_kp_{kn}\hat{A}_k\\
\hat{\mathcal{A}}_{n}^{\perp}=\sum_kq_{kn}\hat{A}_k.
\end{array}
\label{BasisChangeOper}
\end{align}
Expressions in terms of the coefficients $p_{kn}$ and $q_{kn}$ and identical to Eqs.~\eqref{BasisChangeOper} can be obtained for operators $\hat{B}_{out}(\omega)$ and $\hat{C}_{out}(\omega)$, leading to operators $\hat{\mathcal{B}}^{/\!\!/}_{\rm out,n}$, $\hat{\mathcal{B}}^{\perp}_{\rm out,n}$, $\hat{\mathcal{C}}^{/\!\!/}_{\rm out,n}$ and $\hat{\mathcal{C}}^{\perp}_{\rm out,n}$, respectively.

By exploiting the properties~\eqref{psiperp_tran} and~\eqref{psiperp_refl} of functions $\{\psi_{n}^{/\!\!/}(\omega),\psi_{n}^{\perp}(\omega)\}$, the evolution operator can now be written as: 
\begin{align}\label{Efinal}
\hat{\mathcal{U}}&\approx\hat{1}+\theta\sum_{kl}\left[T_{lk}\hat{\mathcal{B}}^{/\!\!/}_{\rm out,l}\left(\hat{\mathcal{A}}^{/\!\!/}_k\right)^\dagger-T_{lk}^*\left(\hat{\mathcal{B}}^{/\!\!/}_{\rm out,l}\right)^\dagger\hat{\mathcal{A}}^{/\!\!/}_k\right]\nonumber\\
&+\theta\sum_{kl}\left[R_{lk}\hat{\mathcal{C}}^{/\!\!/}_{\rm out,l}\left(\hat{\mathcal{A}}^{/\!\!/}_k\right)^\dagger-R_{lk}^*\left(\hat{\mathcal{C}}^{/\!\!/}_{\rm out,l}\right)^\dagger\hat{\mathcal{A}}^{/\!\!/}_k\right]\nonumber\\
&+\theta\sum_k\left[\hat{\mathcal{C}}^{\perp}_{\rm out,k}\left(\hat{\mathcal{A}}^{\perp}_k\right)^\dagger-\left(\hat{\mathcal{C}}^{\perp}_{\rm out,k}\right)^\dagger\hat{\mathcal{A}}^{\perp}_k\right],
\end{align}
with:  
\begin{align}
T_{lk}&=\int\left(\psi_k^{/\!\!/}(\omega)\right)^*t(\omega)\psi_l^{/\!\!/}(\omega)d\omega \nonumber \\
R_{lk}&=\int\left(\psi_k^{/\!\!/}(\omega)\right)^*r(\omega)\psi_l^{/\!\!/}(\omega)d\omega.
\end{align}
The explicit shape of $T_{lk}, R_{lk}$ depends on chosen experimental conditions via $t(\omega)$ and $\psi_l^{/\!\!/}(\omega)$. In Eq.~\eqref{Efinal} operators $\hat{\mathcal{C}}_{out}$ are associated with the spectral components reflected by the filter and will be eventually traced out; accordingly, the most relevant contribution is given by the term $T_{lk}\hat{\mathcal{B}}^{/\!\!/}_{\rm out,l}\left(\hat{\mathcal{A}}^{/\!\!/}_k\right)^\dagger$ and its hermitian conjugate. 

As shown by the explicit expression of $\hat{\mathcal{U}}$, the introduction of the filter BS defines a new basis of modes and associated bosonic operators that permit to describe the whole subtraction process, although, in general, they do not necessarily match the original $\{\psi_k(\omega)\}$ and corresponding $\{\hat{A}_k\}$. Due to such a mismatch, any operation performed in a single detection mode $\psi^{/\!\!/}_k(\omega)$ can have an impact distributed on all original supermodes $\{\psi_k(\omega)\}$ and can even entangle them. This very general result has been experimentally demonstrated in Ref.~\cite{TrepsNG2020} for the mode-selective case. In the special case of non-mode selective operation, expressing $\hat{\mathcal{U}}$ in terms of the parallel modes (as shown in Eq.~\eqref{Efinal}) shows that a photon-counting operation on a single $\psi^{/\!\!/}_k(\omega)$ also affects in a coherent way \emph{all} parallel modes of the heralded state and entangle them too. In the language of single-photon subtraction, measuring a photon-counting event on a single $\psi^{/\!\!/}_k(\omega)$ would correspond to herald the delocalised subtraction of a single-photon on a set of $\{\psi^{/\!\!/}_{k}\}$. It is important to note that since standard photon-counting detectors do not discriminate among signals coming from different $\psi^{/\!\!/}_k(\omega)$, a photon-counting event downstream of the filter eventually leads to a mixed state as it will be discussed in the next paragraph.

\subsection{Heralded single-photon subtracted state} 
The above formalism allows explicitly describing the action of the operator $\hat{\mathcal{U}}$ on any generic protocol input $\ket{\psi}$ at the port $a$ of the subtraction beam-splitter. The results obtained in this section refer to the case of a pure input state $\hat{\rho}_{\rm in}=\ketbra{\psi}$; however, the entire treatment can be easily extended to a mixed input. The states at the input port $b$ of the subtraction beam-splitter and at the input $c$ of the filter BS are assumed to be the vacuum states $\ket{0}$. The mode entering the subtraction beam-splitter will thus be indicated as $\hat{\rho}= \hat{\rho}_{\rm in}\otimes \ketbra{0}$. 
\bigskip

The state after the filter BS but right before the heralding detection stage can be written as  $\hat{\mathcal{U}}~\hat{\rho}~\hat{\mathcal{U}}^{\dagger}$, with $\hat{\mathcal{U}}$ given by Eq.~\eqref{Efinal}. 
As discussed, in general, the photon-counting detector on the heralding path is a bucket one unable to distinguish signals that originate from different parallel modes $\{\psi^{/\!\!/}_{k}(\omega)\}$. Accordingly, the overall heralded state is given by the mixture: 
\begin{align}\label{rhoout}
\hat{\rho}_{\rm out}&=\frac{1}{P}\text{Tr}_{\hat{\mathcal{B}}^{/\!\!/}_{\rm out}} [ \hat{\Pi}_{\hat{\mathcal{B}}^{/\!\!/}_{\rm out}}\hat{\mathcal{U}}~\hat{\rho}~\hat{\mathcal{U}}^{\dagger} ],
\end{align}
where $P=\text{Tr} [ \hat{\Pi}_{\hat{\mathcal{B}}^{/\!\!/}_{\rm out}}\hat{\mathcal{U}}~\hat{\rho}~\hat{\mathcal{U}}^{\dagger} ]$ and $P\cdot \theta^2$ is the protocol success probability. $\text{Tr}_{\hat{\mathcal{B}}^{/\!\!/}_{\rm out}}$ is a partial trace over all detection modes $\{\psi^{/\!\!/}_{n}(\omega) \}$ of $\hat{b}_{\rm out}$. In Eq.~\eqref{rhoout}, the state $\hat{\rho}_{\rm out}$ and the probability $P$ are obtained by performing also the trace over the modes reflected by the filter BS; the explicit indication of the $\text{Tr}_{\hat{\mathcal{C}}_{\rm out}}$ has been omitted to simplify the notation. {Both $\hat{\rho}_{\rm out}$ and $P$ depend on the chosen heralding detector and measurement result via the detector positive operator-valued measure (POVM), $\hat{\Pi}_{\hat{\mathcal{B}}^{/\!\!/}_{\rm out}}$ that accounts, in principle, for limited detection efficiency, photon counting ability, time resolving ability~\cite{Gouz2018Time}.} Here, the operator $\hat{\Pi}_{{\hat{\mathcal{B}}^{/\!\!/}_{\rm out}}}$ reads as: 
\begin{equation}\label{POVM}
\hat{\Pi}_{{\hat{\mathcal{B}}^{/\!\!/}_{\rm out}}} = \sum_n \ketbra{1}_{{\hat{\mathcal{B}}^{/\!\!/}_{\rm out,n}}}
\end{equation}
and is given by the superposition of projectors $\ketbra{1}_{{\hat{\mathcal{B}}^{/\!\!/}_{\rm out,n}}}$, each describing the detection of a single photon in a specific mode $\psi^{/\!\!/}_{n}(\omega)$. 
Without any loss of generality, the bucket detector on the heralding path is assumed here to be able to perfectly project its input on a single-photon state. The impact of poor or no photon number resolving ability on heralded state preparation is discussed in Ref.~\cite{Laurat2012EPJD}. 

By making use of the explicit expression of $\hat{\mathcal{U}}$ reported in Eq.~\eqref{Efinal} as well as of relations~\eqref{BasisChangeOper}, it is easy to obtain the exact expression of the density matrix $\hat{\rho}_{\rm out}$ in terms of the bosonic operators associated with the input supermodes:
\begin{align}\label{rhofinale}
\hat{\rho}_{\rm out}=\frac{1}{P}\sum_{n,k}\gamma_{k,n}\hat{A}_k~\hat{\rho}_{\rm in}~\hat{A}_{n}^{\dagger},
\end{align}
where $P$ can be expressed as:
\begin{align}\label{Pfinale}
P=\sum_{n}\gamma_{n,n \rm} Tr[\hat{A}_n \hat{\rho}_{\rm in} \hat{A}_{n}^{\dagger}].
\end{align}
The coefficients $\gamma_{k,n}$ can be written in a compact form as functions of the filter BS transmission: 
\begin{align}\label{gammank}
\gamma_{k,n}&=\int|t(\omega)|^2\psi_k(\omega)\psi_{n}^*(\omega)d\omega.
\end{align}
Expression~\eqref{rhofinale} describes in a very simple and general form the mixed state heralded by a single-photon subtraction from the initial input state $\hat{\rho}$ in all cases where no mode-selective operation is possible on the heralding path. As anticipated, this corresponds to the extremely common experimental situation in which passive, linear filters are employed on the heralding path~\cite{LvovskyNG2020, branczyk2010optimized, christ2014theory, Furusawa2017CWCats}. The special case of mode-selective single-photon subtraction in a given supermode $\psi_{\bar{n}}(\omega)$ corresponds to $\gamma_{k,n}=\delta_{k,\bar{n}}\delta_{n,\bar{n}}$ in \eqref{rhofinale} and correctly leads to a pure output state $\hat{\rho}_{\rm out}=\frac{1}{P}\hat{A}_{\bar{n}}~\hat{\rho}_{\rm in}~\hat{A}_{\bar{n}}^{\dagger}$~\cite{averchenko2016multimode}.

\subsection{Continuous variable detection of the heralded state}
Non-Gaussian properties of the heralded state $\hat{\rho}_{\rm out}$ can be retrieved by sending it to a homodyne detector (HD) to make it beat with a classical local oscillator beam (LO). This technique is used to fully reconstruct the properties of the produced states via quantum homodyne detection. In experiments, the HD is generally triggered by the photon-counting events detected on the heralding path. As widely discussed in the literature~\cite{LvovskyNG2020, TrepsNG2020}, the homodyne detector acts as a logic AND gate, in which the properties of the LO select the mode that will be actually measured. In particular, by expressing the amplitude of the LO beam as $\alpha_{LO}(\omega)=\sum_k  c_k \psi_k(\omega)$, it is simple to prove that the homodyne detection measures the observable:
\begin{align}\label{HomodyneX}
\hat{X}_{H}(\varphi) = \sum_{k = 0}^{\infty}  c_{k} \hat{X}_{k}(\varphi),
\end{align}
where $\varphi$ is the LO phase  and $\sum_k|c|^2_k=1$. Observable $\hat{X}_{k}(\varphi)$  is the quadrature $\hat{X}_{k}(\varphi) = \dfrac{1}{\sqrt{2}} \left( e^{-i\varphi} \hat{A}_{k} + e^{i\varphi} \hat{A}^{\dag}_{k} \right)$ associated with the supermode $\psi_k(\omega)$. Amplitude and phase quadratures, $\hat{X}$ and $\hat{Y}$, correspond to $\varphi=0,\pi/2$, respectively.  Depending on the chosen spectral profile of the LO, $\alpha_{LO}(\omega)$, it is thus possible to observe the quadratures of a given supermode or of a combination of them. This possibility is extremely relevant when it comes to chosing how to measure the multimode features or to optimise the detection of a chosen mode. By exploiting Eqs.~\eqref{rhofinale} and~\eqref{HomodyneX}, the single mode Wigner function of the state detected by the homodyne detector can be written as: 
\begin{align}\label{Wigner_function_2}
&W_{H}(x, y) = \dfrac{1}{2\pi^2P} \sum_{k, n} \gamma_{k,n} \cdot\nonumber \\ 
&\int e^{i 2 x' y} e^{-i \left( x - x' - \mathbf{c}^{t}\cdot\mathbf{x} \right)z}
 \langle \mathbf{x}\vert \hat{A}_k ~\hat{\rho}_{\rm in}~ \hat{A}^\dag_n \vert \mathbf{x} + 2x' \mathbf{c} \rangle d\mathbf{x}dz dx',
\end{align}
where $\bold{c}^t = ( c_0,..., c_k,...)$ and $\ket{\mathbf{x}} = \ket{ x_0,...,x_k,... }$. The previous expression is obtained by tracing out all modes orthogonal to the one defined by the homodyne via the LO profile~\cite{sasaki2006multimode} as expressed by~\eqref{HomodyneX}.

\section{Heralded generation of Schr\"odinger kitten states}
\label{secCatGen}
\begin{figure}[t] 
  \begin{center}
      \includegraphics[width=1\columnwidth]{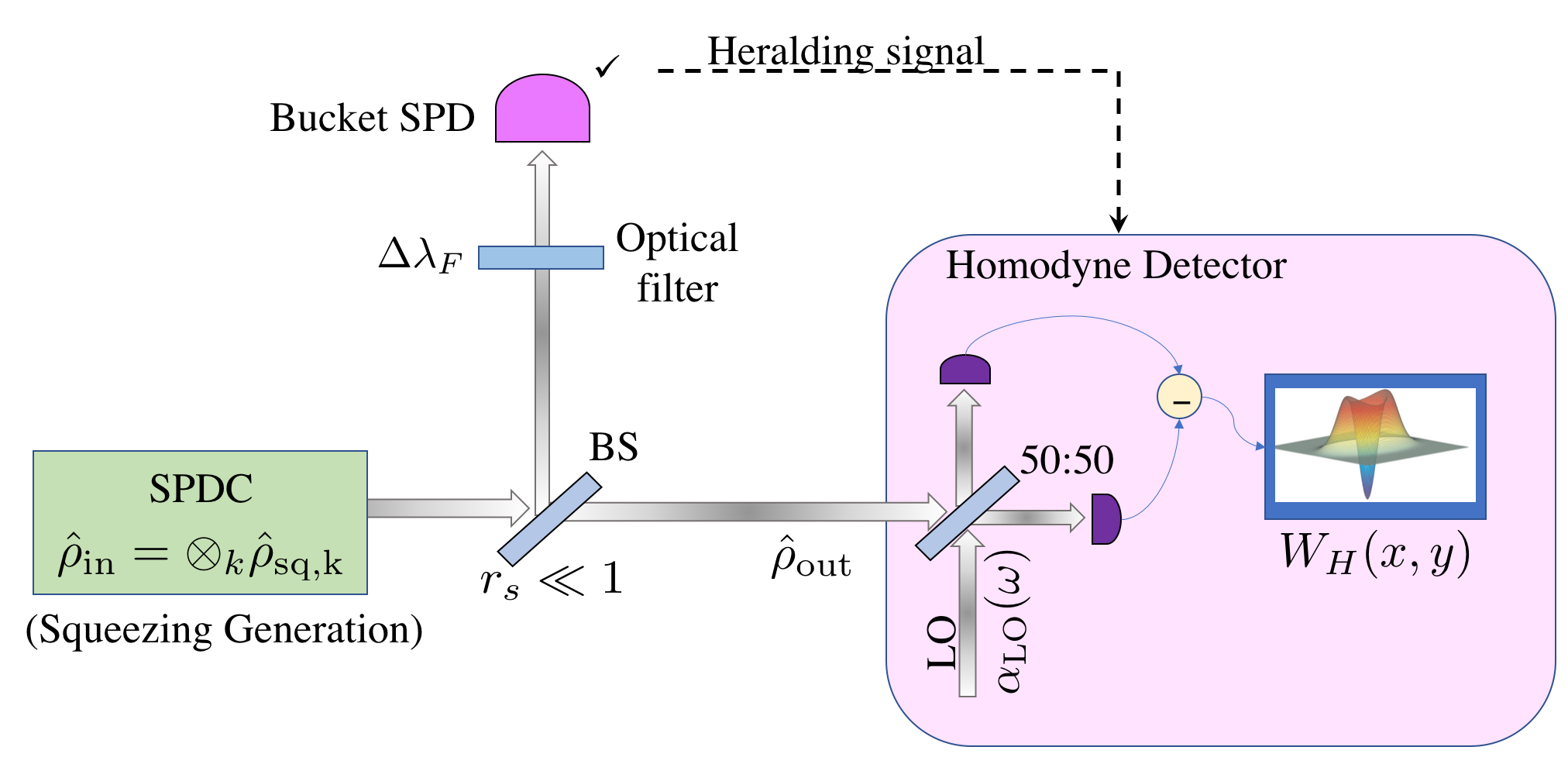}\hspace{-0.5cm}   
 \caption{Heralded preparation and detection of Schr\"odinger kitten states via single photon subtraction from a squeezed state produced by spontaneous parametric down conversion. Depending on the SPDC working condition, a multipartite squeezed state, $\hat{\rho}_{\rm in}$, is produced. An optical passive filter of FWHM $\Delta \lambda_F$ is set before the SPD on the heralding path of the scheme, so as to implement a non mode-selective single-photon subtraction. The SPD detection signal is used to herald the production of the non-Gaussian state $\hat{\rho}_{\rm out}$ as well as to trigger the homodyne detection. The measured state, represented by its Wigner function $W_{H}(x,y)$, strongly depends on the spectral profile $\alpha_{\rm LO}(\omega)$ of the homodyne local oscillator.
}\label{schemaFelino}
\end{center}
\end{figure}

The formalism established in the previous section describes in a general way the heralded state  when single-photon subtraction is performed in a nonmode-selective way on an arbitrary input state $\hat{\rho}$. This section focuses on the interesting case study of the heralded preparation of Schr\"odinger kitten states~\cite{LvovskyNG2020}. The typical heralded state generation and detection scheme is represented in Fig.~\ref{schemaFelino}.

Ideally, the protocol for generating small size Schr\"odinger cats relies on photon subtraction from a single-mode squeezed vacuum state  $\hat{\rho}_{\rm sq,k}=\ketbra{\psi}_{\rm sq,k}$ in a well defined supermode $\psi_k(\omega)$ and exhibiting a low squeezing level~\cite{LvovskyNG2020}. The corresponding state Wigner function is given by: 
\begin{equation}\label{targetammammeta}
W_{T}(x, y) = \dfrac{e^{-(x^{2} / s + s y^{2})}}{\pi} \left[ \dfrac{2}{s} x^{2} + 2 s y^{2} - 1 \right],
\end{equation}
where $s=\dfrac{1 + \mu}{1 - \mu}$ depends on the squeezing parameter $\zeta$ via $\mu=\tanh \zeta$.  In many experimental situations, squeezed vacuum states are obtained by means of spontaneous parametric down conversion of an optical pump in a non linear optical cristal. However, to produce single-mode squeezing, SPDC sources demand to be engineered according to specific characteristics, that cannot be obtained for all wavelengths and non linear materials~\cite{Silberhorn2011SingleModeTWB}. In general, the SPDC process can induce frequency/time correlations among optical components thus leading to a multimode behaviour~\cite{branczyk2010optimized}. In this situation, the squeezed states produced by the SPDC process is rather described by $\hat{\rho}_{\rm in}=\otimes_k\hat{\rho}_{\rm sq,k}$, \emph{i.e.} by the product of independent pure single-mode squeezed vacuum states each in a different supermode $\psi_k(\omega)$. The explicit shape of the supermodes depends on the spectral profile of the SPDC pump beam and phase matching condition; by assuming a pump beam with a Gaussian spectrum, in the weak squeezing regime, $\{\psi_k(\omega)\}$ are given by Hermite functions and have the same global phase factor (leading to real $\gamma_{k,n}$)~\cite{wasilewski2006pulsed}. The squeezing parameter $\zeta_k$ of individual $\hat{\rho}_{\rm sq,k}$ decreases as the supermode index $k$ increases. The effective number of excited modes can be quantified in terms of the Schmidt number~\cite{Christ2011},
\begin{equation}
K=\frac{(\sum_k \zeta_k^2)^2}{\sum_k \zeta_k^4}.
\end{equation}
The Schmidt number $K$ is equal to 1 in the single-mode case and it increases with the number of supermodes with non-negligible excitation. In experiments, it depends on the SPDC pump spectral width as well as on the non linear medium opto-geometrical properties at the chosen working wavelengths~\cite{Christ2011}.
The expression of the output state provided by~\eqref{rhofinale} can be applied to the general case of a single-photon subtraction at the output of a multimode SPDC process. Correspondingly, the Wigner function of the heralded state, as measured by the homodyne detector, can be expressed in terms of the mean photon number of squeezed states, $n_k = \sinh^{2} (\zeta_{k})$, and of $\mu_k=\tanh \zeta_k$, as:
\begin{align}\label{Multimode_filtered_Wigner_function}
&W_{H}(x, y) = \dfrac{e^{-\left(\frac{x^{2}}{2\sigma_{x}^{2}} + \frac{y^{2}}{ 2\sigma_{y}^{2}}\right)}}{2\pi \sigma_x \sigma_y P} \nonumber \\
&\left[ \sum_k \gamma_{k,k} n_k +\dfrac{1}{2\sigma_x^2} \left( \dfrac{x^2}{\sigma_x^2} -1 \right) \sum_{k, n} \gamma_{k,n} \dfrac{\mu_k \mu_n c_k c_n}{(1 - \mu_k) (1 - \mu_n)} \right. \nonumber\\
&+ \left. \dfrac{1}{2\sigma_y^2} \left( \dfrac{y^2}{\sigma_y^2} -1 \right) \sum_{k, n} \gamma_{k,n} \dfrac{\mu_k \mu_n c_k  c_n}{(1 + \mu_k) (1 + \mu_n)} \right],
\end{align}
with 
\begin{equation}
\sigma_x^2 = \dfrac{1}{2} \sum_{n}  c_n^2 \dfrac{1 + \mu_n}{1 - \mu_n}\quad \text{and}\quad\sigma_y^2 = \dfrac{1}{2} \sum_{n}  c_n^2 \dfrac{1 - \mu_n}{1 + \mu_n}.
\end{equation}
Thanks to expression~\eqref{Multimode_filtered_Wigner_function}, it is possible to explicitly evaluate the impact on the measured state of both the filter profile, via the $\{\gamma_{k,n}\}$, and of chosen the LO shape via the $\{c_n\}$. The multimode features of the protocol input state are taken into account via $n_k$ and $\mu_k$. 
 
\subsection{Measured heralded state without filtering on the heralding path}
The case of heralded states obtained when no filter is used on the heralding path provides a good intuition of the relevance of the filtering stage applied in the protocol. The absence of filter is described by $t(\omega)=1$ for all $\omega$, \emph{i.e.},  according to Eq.~\eqref{gammank}, by $\gamma_{k,n}=\delta_{k,n}$. Trivially, in this case, parallel modes $\{\psi_{k}^{/\!\!/}(\omega)\}$ exactly correspond to $\{\psi_k(\omega)\}$ (see Eq.~\eqref{othonormalization}). As indicated by Eq.~\eqref{rhofinale}, the protocol output state is a perfect mixture of photon subtracted states of the kind $\hat{A}_n\hat{\rho}_{\rm in}\hat{A}_{n}^{\dagger}$. Correspondingly, the Wigner function of the detected state can be expressed as:
\begin{equation}\label{Wigner function}
W_{H,\text{NF}}(x, y) = \sum_{k} p_{k} \left[  c_k^{2} W_{k}^{(1)}(x, y) + (1 -  c_k^{2}) W_{\rm SVS}(x, y) \right],
\end{equation}
with $p_k = \frac{n_k}{\sum_{l} n_{l}}$ the probability of subtracting a photon from the mode $k$. 
The first term of the previous expression represents the Wigner function of a single-photon subtracted squeezed state in supermode $\psi_k(\omega)$ as seen by the HD: 
\begin{align}\label{Wigner_noFilter_Nono}
W_{k}^{(1)}(x, y) = \dfrac{e^{-\left(\frac{x^{2}}{2\sigma_{x}^{2}} + \frac{y^{2}}{ 2\sigma_{y}^{2}}\right)}}{2 \pi \sigma_{x} \sigma_{y}}
\left[ \dfrac{(1 + \mu_{k})}{2\sigma_{x}^{2}(1 - \mu_{k})} \left(\frac{x^{2}}{\sigma_{x}^{2}}-1\right)\right.\nonumber\\
\left. + \dfrac{(1 - \mu_{k})}{2\sigma_{y}^{2}(1 + \mu_{k})} \left(\frac{y^{2}}{\sigma_{y}^{2}} -1\right)+1 \right].
\end{align}
This state correctly corresponds to the target state of Eq.~\eqref{targetammammeta} when the homodyne LO is in the supermode $\psi_k(\omega)$. In general cases, the effect of imperfect mode matching with the LO is included in $\sigma_{x}$ and $\sigma_{y}$. \\
The other term corresponds to squeezed vacuum contributions as detected by the homodyne detector.
\begin{equation}\label{Wigner_squeezed_vacuum}
W_{\rm SVS}(x, y) = \dfrac{e^{-(x^{2} / 2\sigma_{x}^{2} + y^{2} / 2\sigma_{y}^{2})}}{2 \pi \sigma_{x} \sigma_{y}}
\end{equation}
Equation~\eqref{Wigner function} expresses the fact that, in absence of filter, photon-counting events cannot be associated to a specific supermode (nor to any subset of them); as a consequence, the state measured by the HD only occasionally corresponds to the desired non-Gaussian one. A physical intuition of this idea is provided by the case of a LO in a specific supermode $\alpha_{\rm LO}(\omega)=\psi_{n}(\omega)$, \emph{i.e.}  $|c_k|^{2} = \delta_{k,n}$. In this condition:  
\begin{equation}\label{WnofilterLOmatched}
W_{H, \text{NF}}(x, y)\vert_{ |c_n| ^2= 1} = p_{n} W_{n}^{(1)}(x, y) +\sum_{k\neq n} p_{k} W_{\rm SVS}(x, y).
\end{equation}
The measured heralded state is equal to $W_{n}^{(1)}(x, y)$ only with a probability $p_{n}$, corresponding to the probability of subtracting a photon in the LO mode $\psi_n(\omega)$. All remaining cases lead to the detection of squeezed vacuum with a probability $\sum_{k\neq n} p_{k}=1-p_n$. 
 
State non-classicality can be evaluated using several metrics. For a pure state, non-Gaussianity implies a Wigner function that is negative in a given region in phase space~\cite{hudson1974wigner}. However, mixed states with non-Gaussian features can still exhibit a positive Wigner function. In these conditions, specific criteria can be adopted to quantify the non-Gaussianity of a quantum state~\cite{genoni2013detecting}. At the same time, to discuss \emph{optimally accessible} features of heralded states, it is pertinent to focus on the Wigner function negativity, as it represents a stricter metric~\cite{filip2011detecting}. Note that, in what concerns all results that will be presented in the following sections, the Wigner function negativity and its non-Gaussianity share the same qualitative behaviour. The negativity is defined as the volume of the negative part of the Wigner function \cite{kenfack_2004_negativity} and it can be expressed as:
\begin{equation}
N_g = \dfrac{1}{2} \left[ \iint \vert W_H(x, y) \vert dx dy - 1 \right].
\end{equation}
The negativity for the ideal case of a single-photon subtraction on a single-mode squeezed vacuum state is computed to be $N_g = 2 e^{-1/2} - 1 \approx 0.213$.


\begin{figure} 
\begin{tabular}{c}
 \includegraphics[width=\columnwidth]{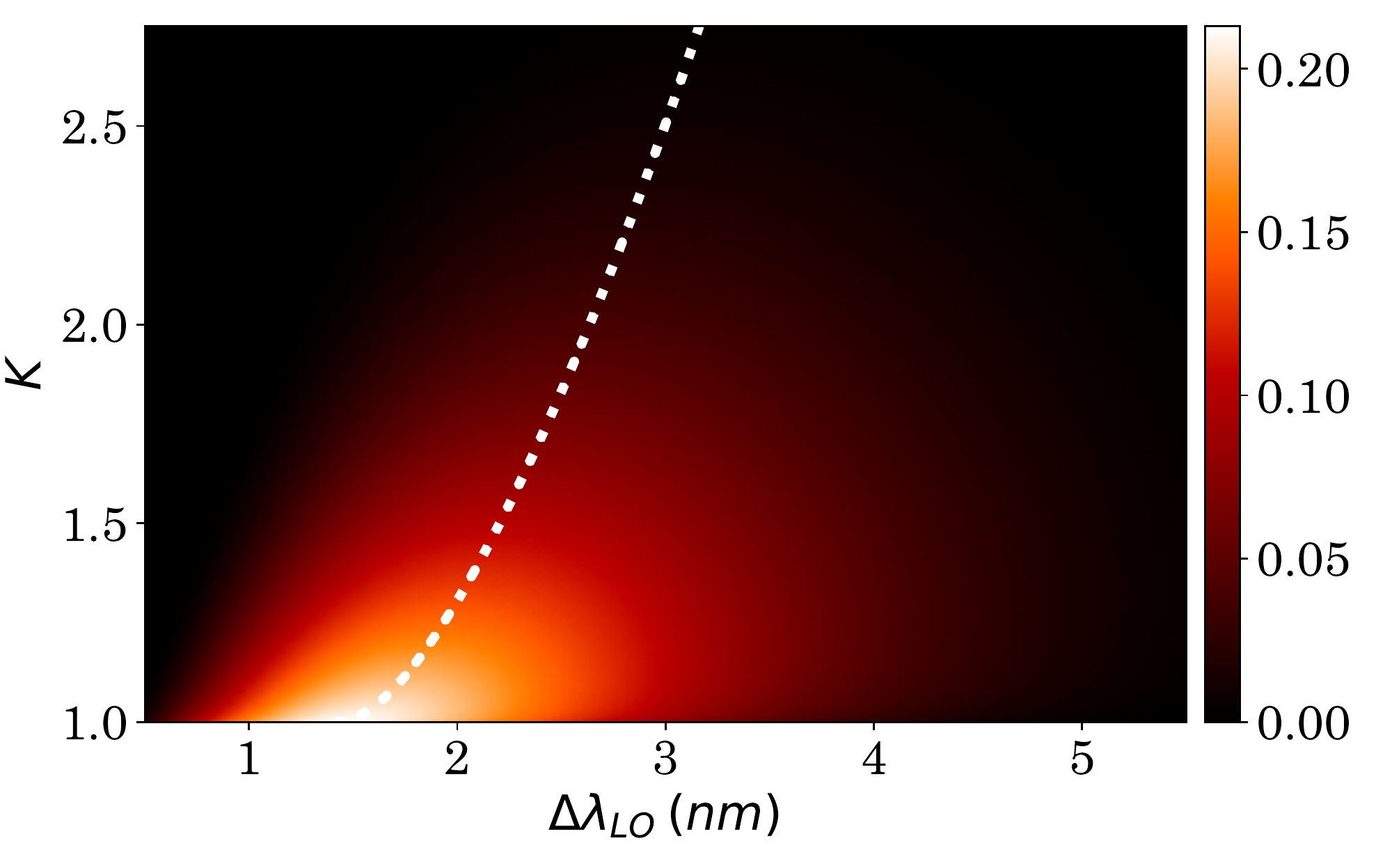} \\ 
 \includegraphics[width=\columnwidth]{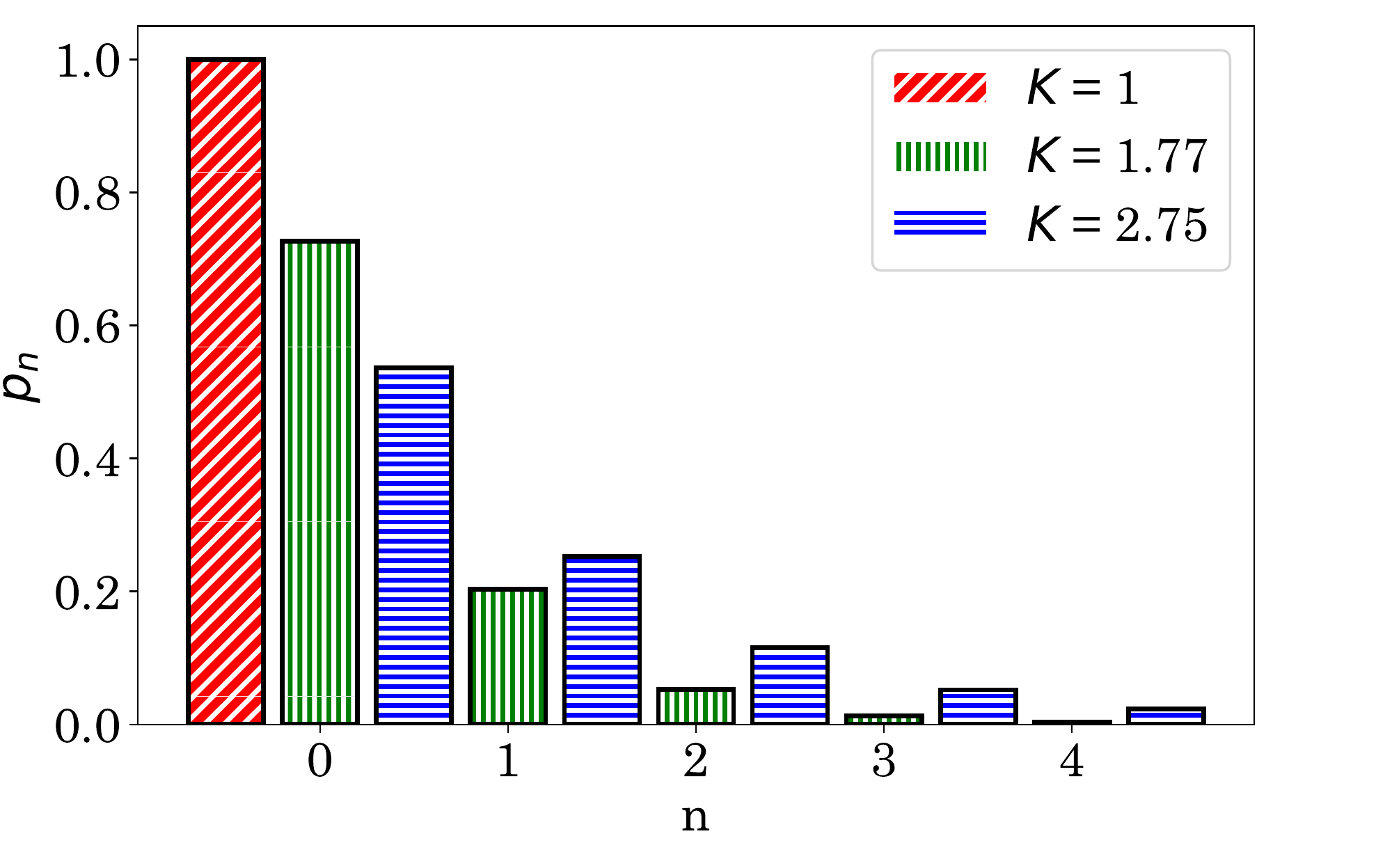} \\ 
 \end{tabular}  
\caption{(Color online) Properties of the measured heralded state $W_{H,\text{NF}}(x, y)$ in the absence of filter on the heralding path. \emph{Top}: Wigner function negativity $N_g$ of $W_{H,\text{NF}}(x, y)$ as a function of the local oscillator spectral bandwidth and the multimode character of the SPDC source providing the input state $\hat{\rho}_{\rm in}$ via $K$; the LO spectrum is assumed to be Gaussian with a FWHM $\Delta\lambda_{LO}$. The white dotted curve indicates the case of a local oscillator perfectly matched with the first SPDC supermode (i.e. $|c_0|^{2}=1$). \emph{Bottom}: Probability $p_n$ of detecting an heralding photon from the supermode $\psi_n(\omega)$ for different SPDC working condition. Probabilities $p_n$ are directly proportional to the excitation of the squeezed supermodes $\{\psi_n(\omega)\}$. In all the simulations, the bandwidth of the SPDC pump beam is set to 0.5\,nm. $K$ depends on the SPDC pump spectral width as well as on the the non linear medium opto-geometrical parameters at the chosen working wavelength~\cite{Christ2011}.}\label{fig_Schmidt}
\end{figure}

Fig.~\ref{fig_Schmidt}-\emph{top} shows $N_g$ for the detected heralded state as a function of the local oscillator spectral width and of the Schmidt number $K$. To comply with many experimental cases, in which the LO beam is directly provided by the output of a pulsed laser, the LO profile, $\alpha_{LO}(\omega)$, is assumed to be Gaussian with FWHM of $\Delta\lambda_{LO}$. As shown by the figure, optimal $N_g$ values are obtained for low $K\approx 1$ and when the Gaussian LO profile $\alpha_{LO}(\omega)$ is chosen so as to match exactly the first SPDC supermode, \emph{i.e.}  $\alpha_{LO}(\omega)=\psi_0(\omega)$ and $|c_0|^{2}=1$. This case is represented in the figure by the white dotted line. When the number of supermodes at the SPDC output increases ($K >1$), $N_g$ rapidly degrades. This behaviour can be understood by observing Fig.~\ref{fig_Schmidt}-\emph{bottom}, that represents the probability $p_n$ of subtracting a photon in the supermode $n$; as it can be seen, the probability $p_0$ is generally higher than $p_{n>0}$ and the single-photon subtraction mostly takes place on the first supermode ($n=0$). Accordingly, choosing the LO profile identical to the first supermode ($\alpha_{LO}(\omega)=\psi_0(\omega)$) corresponds to optimising, in Eq.~\eqref{WnofilterLOmatched}, the detection of the non-Gaussian state $W_{n}^{(1)}(x, y)$ that shows a negative Wigner function. Note that, to measure high negativities, contributions with $p_{n>0}$, that lead in Eq.~\eqref{WnofilterLOmatched} to the detection of squeezed vacuum, should be minimised. This condition is easily respected when $K\approx$1. However, as soon as the Schmidt number increases, the distribution of $p_n$ becomes flatter, thus explaining why, for high $K$ values, $N_g$ progressively decreases along the white dotted line.

\subsection{Rectangular or Gaussian spectral filters in the heralding path}

In experiments, the use of narrowband filters on the heralding path represents a widely reported strategy to improve the quality of single-photon subtracted states~\cite{LvovskyNG2020}. At the same time, to our knowledge, no theoretical and quantitative description of the filter impact has been provided so far. The formalism that has been introduced in this work can be conveniently use to illustrate the limits and the advantages of different filter strategies. As discussed in the previous sections, the actual spectral profile of the function $t(\omega)$ determines the shape of the filter supermodes $\{\psi_{k}^{/\!\!/}(\omega),\psi_{k}^{\perp}(\omega)\}$  and, via Eq.~\eqref{gammank}, the explicit shape of the coefficients $\gamma_{k,n}$ to be included in $W_H(x,y)$ of Eq.~\eqref{Multimode_filtered_Wigner_function}. Accordingly, different filter profiles and FWHM can have a strong effect on the heralded states as well as on the LO profile that must be chosen to optimise the detection of non-Gaussian features.  
From the experimental side, different filter spectral profiles can be chosen depending on the specific situation. Plug-and-play filters with a rectangular profile, transmitting light only within a certain bandwidth are widely used in guided-wave experiments, where broadband SPDC emission is generated in single-pass configuration in non linear waveguides~\cite{Silberhorn2011SingleModeTWB, Mondain2019}. This scenario covers dense wavelength multiplexers (DWM) and Bragg filters and it is quite common when quantum optical states are produced at classical telecommunication wavelengths so as to be compatible with low loss fibre networks. Alternatively, filters with $t(\omega)$ described by continuous and strictly non-null functions, such as Gaussian or Lorentzian, can be used. This is the case, for instance, of bulk configurations in which DWM or Bragg filters with desired characteristics in terms of transmission or spectral width are not available and optical cavities are instead employed on the heralding path~\cite{Furusawa2017CWCats}. In the following, the cases of a rectangular filter and of a Gaussian-shaped $t(\omega)$ will be considered and compared. The explicit expressions of $\gamma_{k,n}$ for these cases are reported in Appendix~\ref{Appendice3}. As for the previous section, a Gaussian LO profile $\alpha_{LO}(\omega)$ will be considered.

\begin{figure}
\begin{tabular}{c}
 \includegraphics[width=1\columnwidth]{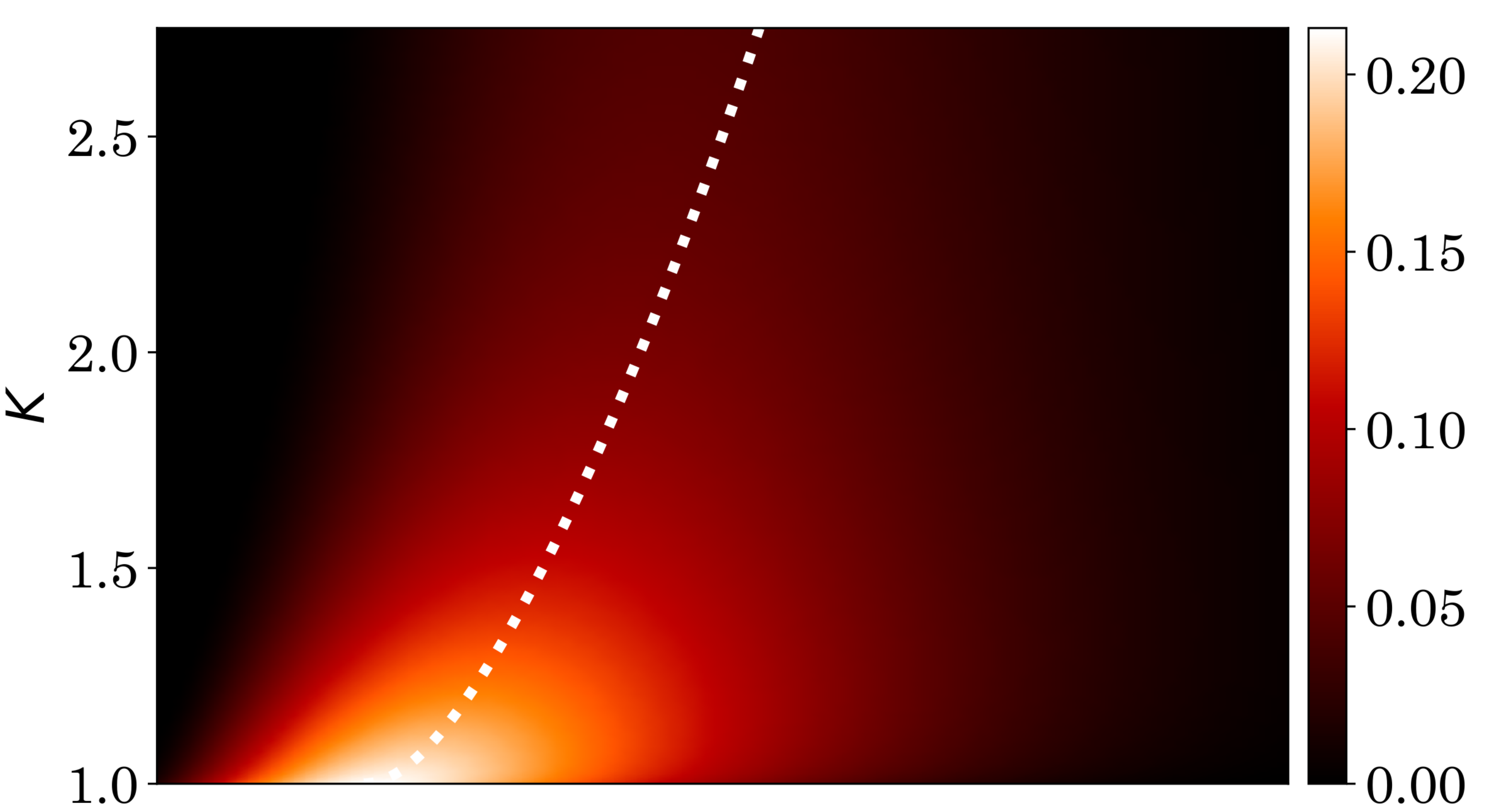} \\ 
 \includegraphics[width=1\columnwidth]{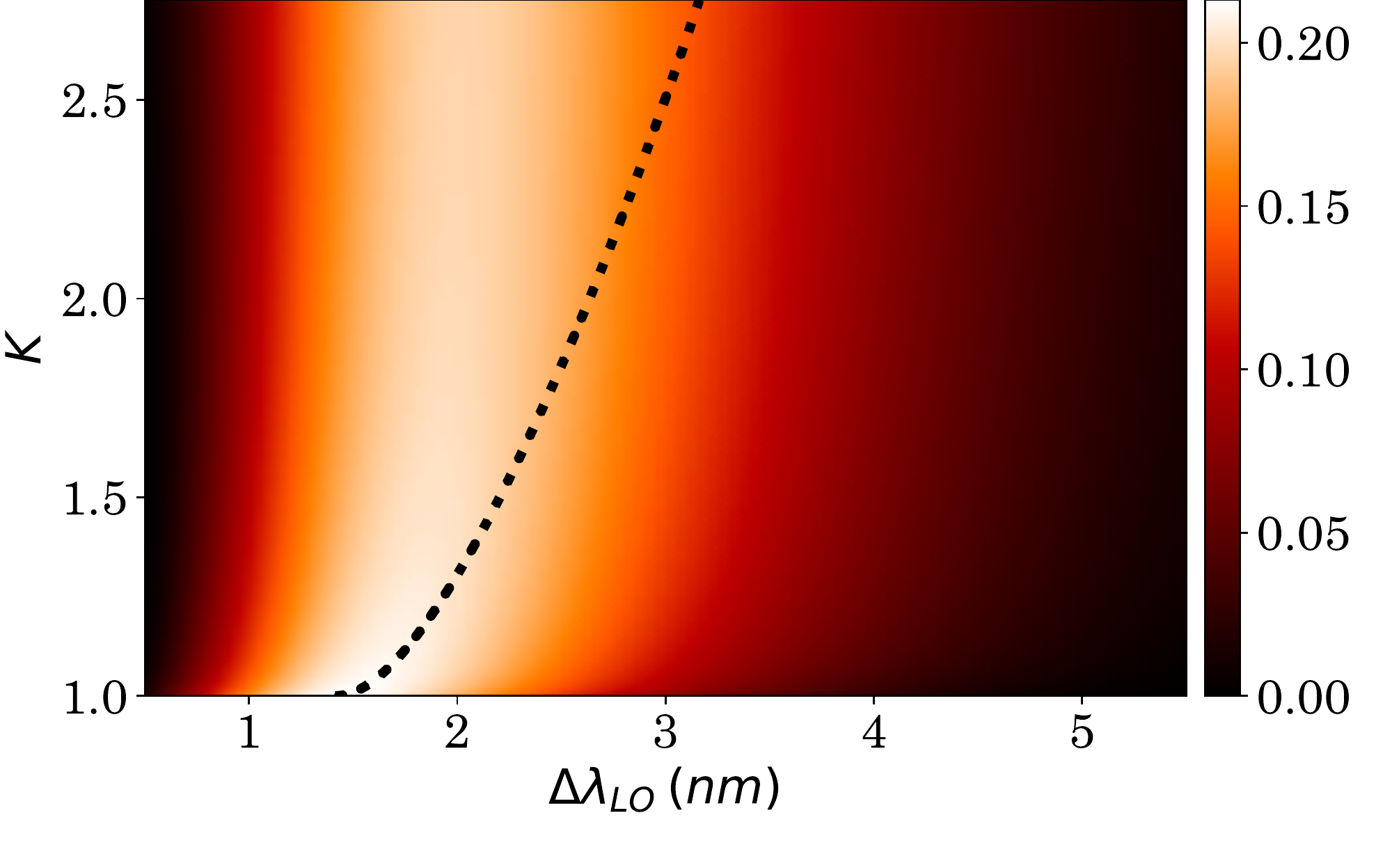} \\ 
\end{tabular} 
 \caption{(Color online) Negativities $N_g$ of the measured heralded Wigner function $W_{H}(x, y) $ in the case of a rectangular filter on the heralding path expressed as functions of the local oscillator spectral bandwidth and Schmidt number $K$. Two filter bandwidths are considered: \emph{top} 5\,nm ($\approx$600\,GHz) FWHM and \emph{bottom} 1\,nm ($\approx$100\,GHz)) FWHM.  In both plots, the dotted curve indicates the case of a local oscillator perfectly matching the first SPDC supermode (i.e. $|c_0|^{2}=1$).}\label{Particular_Filter}
\end{figure}
\bigskip

Fig.~\ref{Particular_Filter} gives the Wigner function negativity for the detected heralded state in the case of a rectangular spectral filter in the heralding path. The first  $\{\psi_{n}^{/\!\!/}(\omega)\}$ and $\{\psi_{n}^{\perp}(\omega)\}$ for this kind of $t(\omega)$ are reported in Fig.~\ref{ModiParalleliOrtogonali} for the case of a filter FWHM of 5\,nm.
 Two different filter bandwidths are considered (5\,nm and 1\,nm). As a general consideration, for both FWHMs, $N_g$ decreases when the number of excited supermodes increases ($K>1$). In the case of a wide filter (Fig.~\ref{Particular_Filter}-\emph{top}), the degradation is quite fast, with a behaviour similar to the one discussed for the case of no filter in the previous section. The optimal LO is still very close to the case of $\alpha_{LO}(\omega)=\psi_0(\omega)$, matching the first SPDC supermode  (\emph{i.e.} $|c_0|^2=1$, see the white dotted curve). The situation dramatically changes when the width of the filter is reduced (Fig.~\ref{Particular_Filter}-\emph{bottom}). In this case, the protocol performances are much robuster against the multimode character of the SPDC source and high negativity values can be obtained also for higher $K>1$, provided the FWHM $\Delta \lambda_{LO}$ for the Gaussian LO is suitably chosen. Remarkably, the optimal $\alpha_{LO}(\omega)$ profile does not match any of the supermodes $\{\psi_k(\omega)\}$, nor any of the filter-defined functions $\{\psi_{n}^{/\!\!/}(\omega)\}$ and $\{\psi_{n}^{\perp}(\omega)\}$ (see Fig.~\ref{ModiParalleliOrtogonali}). Its FWHM $\Delta \lambda_{LO}$ can be easily computed numerically. The possibility of using a Gaussian LO in combination with a narrowband filter confirms qualitative experimental observations~\cite{ourjoumtsev2006, neergaard2006, wakui2007} and opens the way to simple realisations, free from spectral shaping stages on the local oscillator path and mode-selective single-photon subtraction. 


\begin{figure}[hbtp]
\begin{tabular}{c}
\includegraphics[width=\columnwidth]{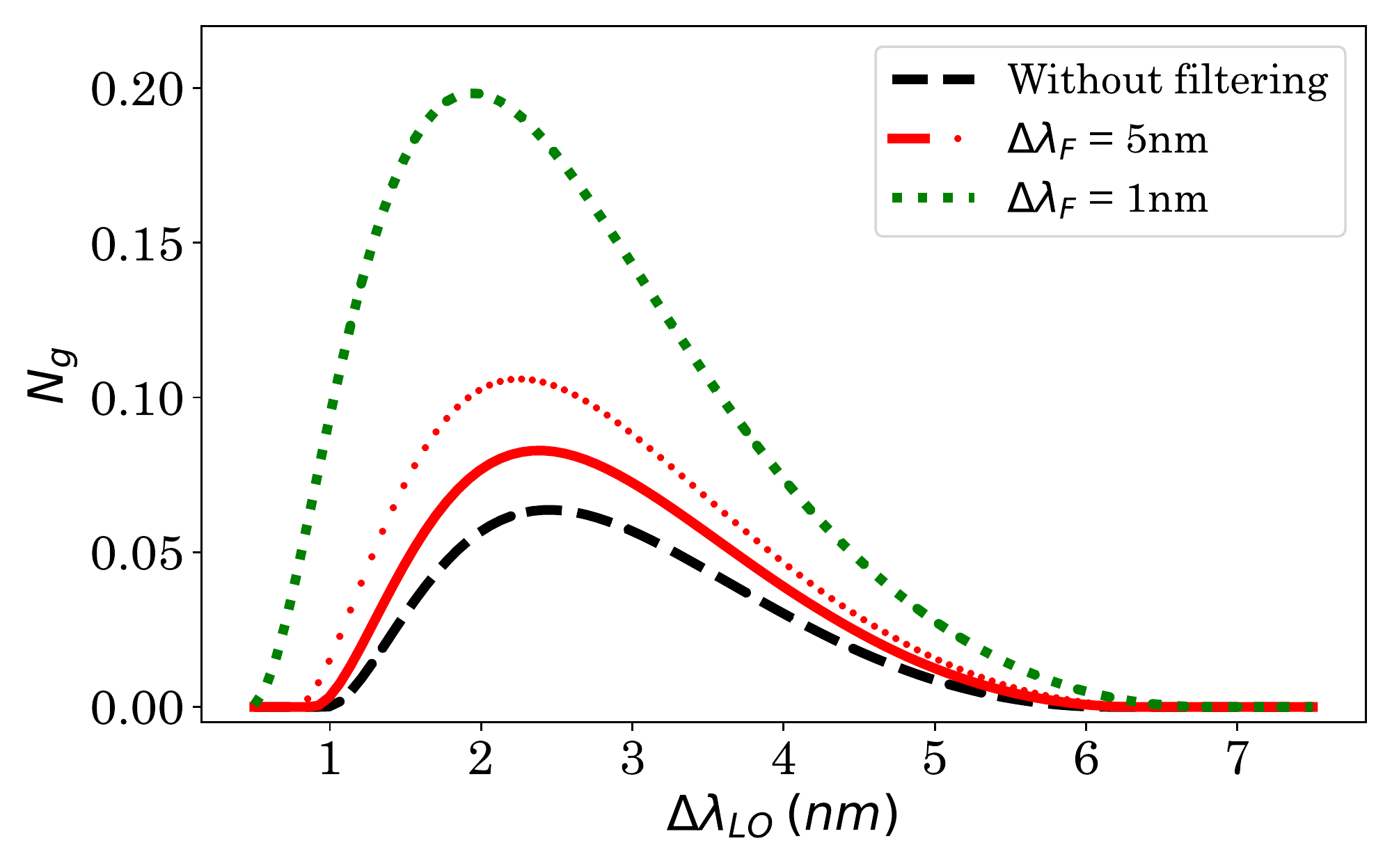} \\ 
\includegraphics[width=\columnwidth]{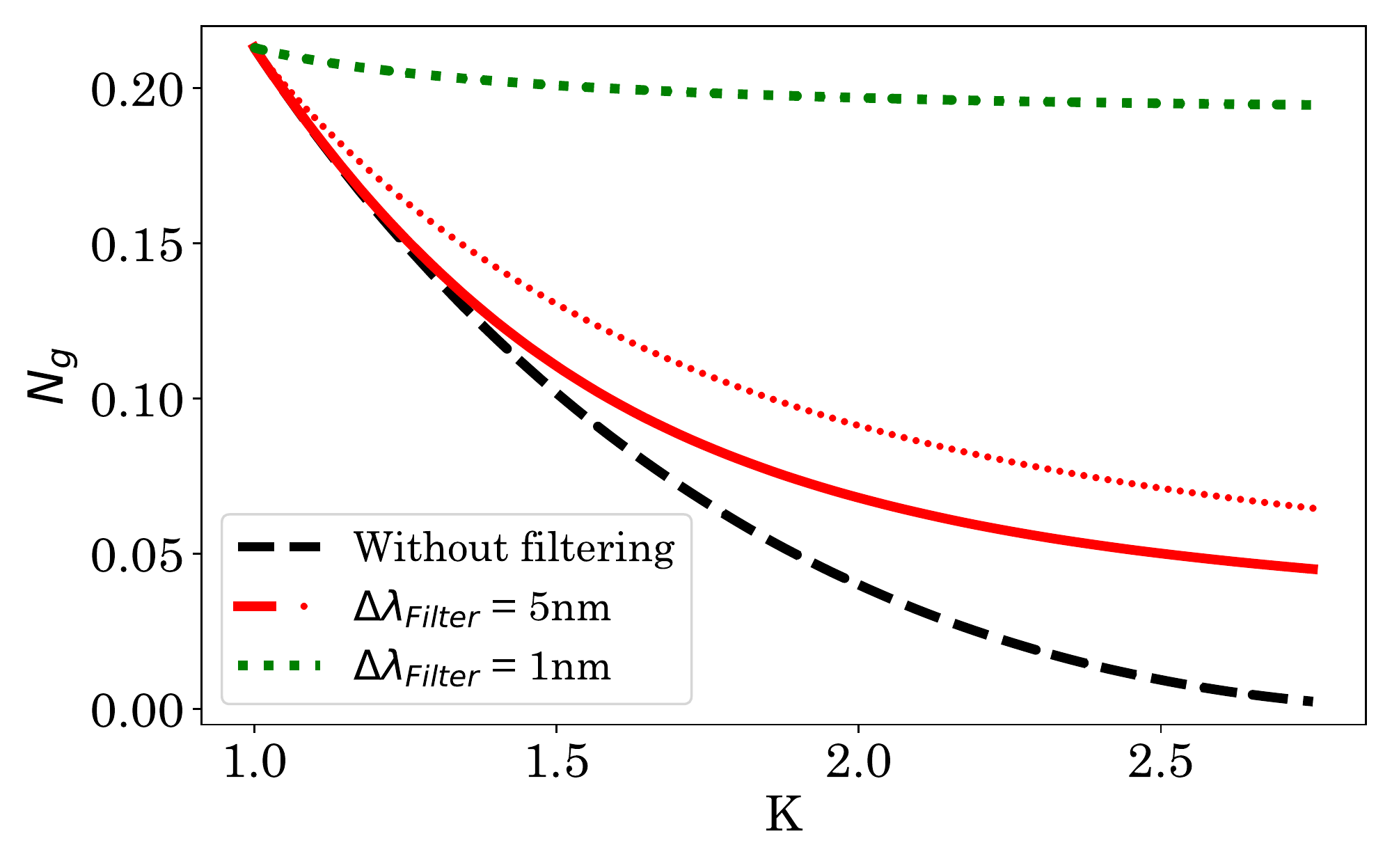} \\ 
\end{tabular} 
\caption{(Color online) \emph{Top}: Negativity of the Wigner function as a function of the FWHM of the LO spectral amplitude $\alpha(\omega)$ in the case where $K$ = 1.77. \emph{Bottom}: Negativity of the Wigner function as a function of the Schmidt number $K$ when imposing an optimal $\Delta \lambda_{LO}$. For both figures the pump SPDC has a spectral bandwidth of 0.5\,nm. Curves refer to a single-photon subtraction scheme. Dashed (black): without spectral filter in the heralding path; solid (red): with 5\,nm ($\approx$\,600\,GHz) rectangular spectral filter in the heralding path; points (red): with 5\,nm ($\approx$\,600\,GHz) Gaussian spectral filter in the heralding path; dotted (green): with 1\,nm ($\approx$\,100\,GHz) spectral filter in the heralding path (the Gaussian and rectangular shapes are equivalent in this case).}\label{Comparison}
\end{figure}
The qualitative behaviour discussed for the case of a rectangular filter is found also in the case of a Gaussian $t(\omega)$. In this case, only the $\{\psi_{n}^{/\!\!/}(\omega)\}$ are non-null. Their profile, similar but not exactly matching the one of the original supermodes, is represented in Fig.~\ref{ModiParalleliOrtogonali}-(a). As for the rectangular filter, reducing the FWHM of the Gaussian $t(\omega)$ has the effect of making the protocol robuster against higher $K$ and makes it easy to identify an optimal LO bandwidth $\Delta \lambda_{LO}$. A comparison among the different filter shapes is shown in Fig.~\ref{Comparison} in the case of an SPDC pump with spectral width of 0.5\,nm. Fig.~\ref{Comparison}-\emph{top} represents the negativity of the Wigner function in the case of no spectral filtering in the heralding path (dashed black), as well as of 5\,nm (red solid and red points) and 1\,nm (dotted green) filter bandwidths for $K$= 1.77. The red solid line represents rectangular spectral filters whereas the red points represent Gaussian spectral filters. As expected, the narrowband spectral filters improve the negativity of the homodyne detected state. Also, the optimal LO bandwidth is slightly shifted to lower values when the filter FWHM decreases. The narrower is the filter, the smaller is the difference between the behaviour obtained with a rectangular or a Gaussian filter profile (dotted green matches solid green curve). As shown in Fig.~\ref{Comparison}-\emph{bottom}, similar results can be found for different values of $K >1$, where optimal measured negativities (\emph{i.e.} corresponding to the optimal LO profiles) are reported as a function of $K$ and for different filter widths. It is relevant to stress that the pertinence of narrowband filters in heralding schemes is equally confirmed in the context of low squeezing and heralded single photon sources, where pure single photon state can be retrieved, even from a highly multimode squeezed states, when delta-like $t(\omega)$ are employed. This case is discussed in detail in Appendix~\ref{Appendice4}.
\bigskip

\subsection{Application to a practical experimental situation}

\begin{figure}[hbtp]
\begin{tabular}{c}
\includegraphics[width=\columnwidth]{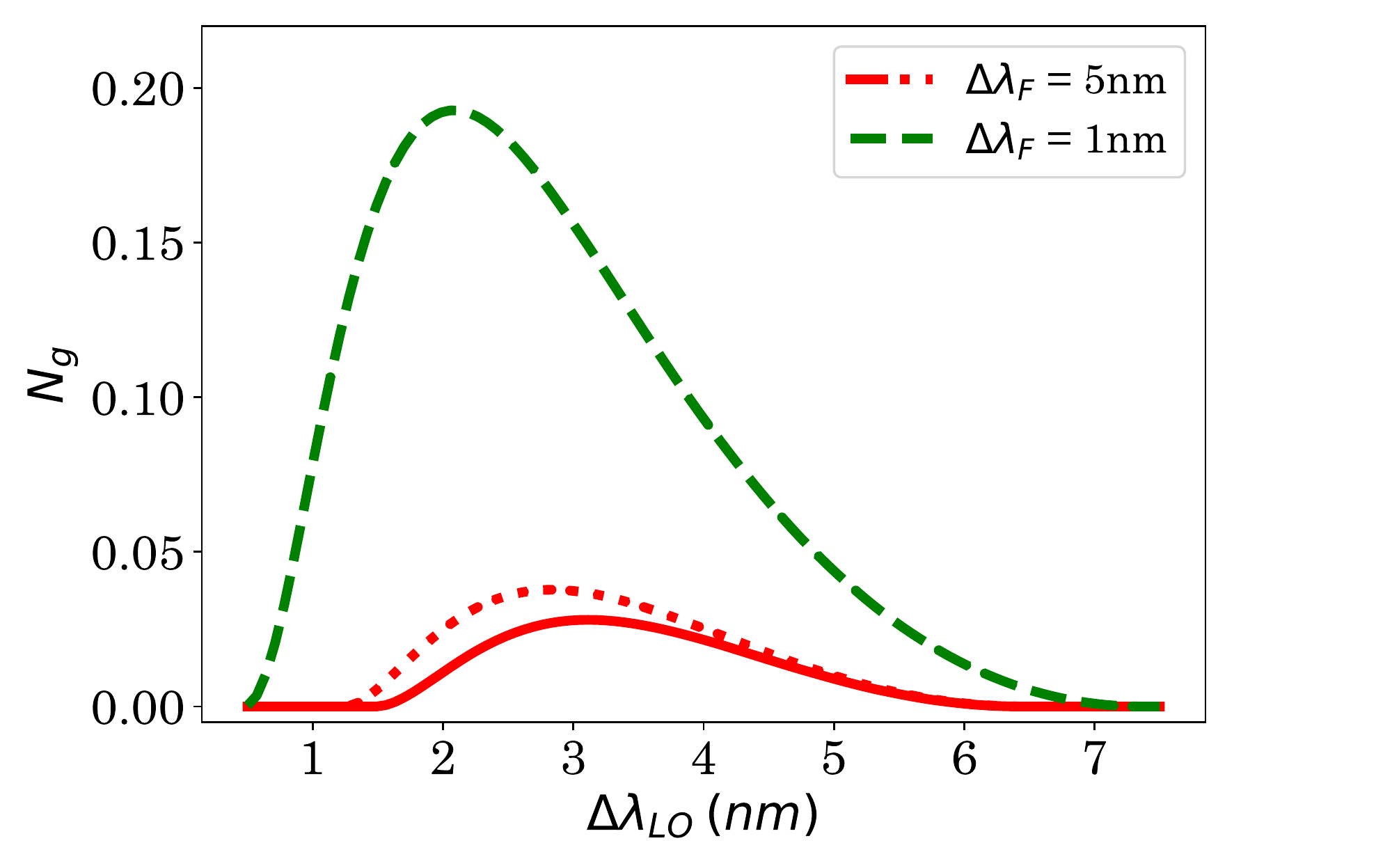} \\ 
\includegraphics[width=\columnwidth]{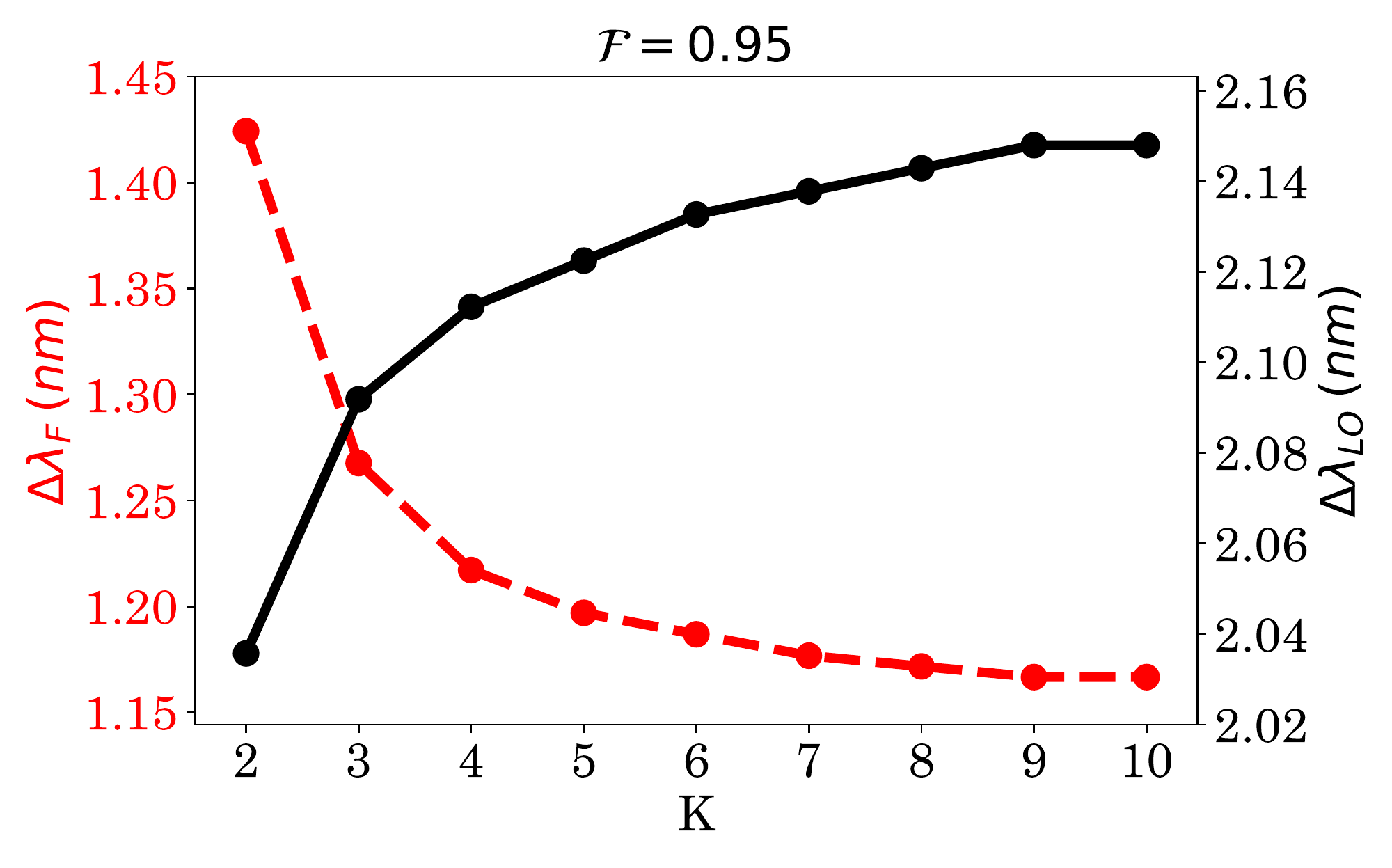} \\ 
\end{tabular} 

\caption{(Color online) \emph{Top}: Negativity of the Wigner function as function of the LO spectral width, for $K\approx$ 9 as for frequency degenerate SPDC around 1560\,nm, pumped with optical pulses at 780\,nm with a width of 0.5\,nm. The SPDC crystal is taken to be 10\,mm-long. Curves refer to a single-photon subtraction scheme. Solid (red): with 5\,nm rectangular spectral filter in the heralding path; dotted (red): with 5\,nm Gaussian spectral filter in the heralding path; dashed (green) with 1\,nm spectral filter in the heralding path (the Gaussian and rectangular shapes are equivalent in this case). \emph{Bottom}: Optimal $\Delta \lambda _{\rm LO}$ (solid black line) and $\Delta \lambda_{F}$ (dashed red line) required to ensure, for a given $K$, a fidelity with the target state of 0.95.}\label{CasoReale}
\end{figure}

Discussed results provide a powerful tool to model single-photon subtraction in a multimode context and to optimise experimental conditions for the generation and detection of high quality non-Gaussian states. As for an example, in what follows, the developed treatment is applied to the practical situation of Schr\"odinger kitten states obtained from single-photon subtraction from squeezed light emitted by a realistic source. A particularly interesting case is that of a SPDC process occurring in integrated photonic circuits on Lithium Niobate (LN), where on-chip generation and active manipulation of optical quantum states at telecom wavelengths can be provided by the association of the material non linear optical- and electro-optical properties~\cite{Mondain2019, Lenzini}. At the same time, in pulsed regime, LN dispersion properties prevent from working in single mode conditions and generated squeezed light can exhibit a high number of modes ($K>1$). Fig.~\ref{CasoReale} considers a typical case of multimode light generated around 1560\,nm via type-0 SPDC in a 10\,mm-long LN crystal pumped by a pulsed beam at 780\,nm with a spectral width of 0.5\,nm. In such conditions, the Schmidt number can be calculated to be $K\approx9$~\cite{wasilewski2006pulsed}. Fig.~\ref{CasoReale}-\emph{top} shows the obtained $N_g$ for different filtering and LO choices. As it can be seen, for a filter of 1\,nm,  almost optimal $N_g$ values can be reached for experimentally accessible Gaussian LO profiles with $\Delta \lambda_{LO}\approx2$\,nm. This corresponds to optical pulses in picosecond regime, that can be easily employed in guided-wave configurations, with the possibility of limiting parasitic non linear optical effects and chromatic dispersion in optical fibres. For a subtraction BS reflectivity $r_s^2\approx$\,0.05, the corresponding protocol success probability, {$P\cdot \theta^2$}, is on the order of 0.01, in agreement with typical values reported in the literature~\cite{LvovskyNG2020, ourjoumtsev2006,neergaard2006, wakui2007}. Higher probabilities can in principle be obtained with wider filters at the price of a reduced $N_g$. {The protocol heralding rate is measured in\,Hz and it is the product of $P\cdot \theta^2$ with the experiment clock, \emph{i.e.} the pulse repetition rate (PRR) of the laser pumping the SPDC stage used to generate the squeezing. Its value corresponds to the count rate at the output of the DV heralding detector that triggers the homodyne measurement. For a fixed $P\cdot \theta^2$, the ultimate limit to the protocol heralding rate is given by the bandwidth of the slower between the DV and CV detector. From one side the PRR must be chosen so as to guarantee a DV detection rate below the photon counter saturation level. From the other, even if this condition is satisfied, the heralding rate must also be kept below the homodyne electronic bandwidth, so as to avoid triggering the acquisition at a speed non-compatible with the detector time resolution~\cite{Matthews2021}. Commercial single photon detectors can safely work with count rate below 10\,MHz, corresponding, in the case of $P\cdot \theta^2\approx$\,0.01, to a maximum PPR of $\approx$1\,GHz. These values are fully compatible with commercial homodyne systems that have typical bandwidths of the order of 100\,MHz. Ultra-fast heralding rates above 100\,MHz (with PRR higher that 10\,GHz) could be obtained by multiplexing the DV detectors and employing research-grade homodyne with bandwidth in the GHz regime~\cite{Matthews2021}.} 

{Eventually, note that results observed with $N_g$ are confirmed by }the fidelity between the measured heralded state and the target one of Eq.~\eqref{targetammammeta} (see Appendix~\ref{appendice2} for the explicit calculation):
\begin{equation}
\mathcal{F} = 2\pi \iint W_{H}(x, y)W_{T}(x, y) dx dy.
\end{equation}
In the simulations, the squeezing parameter of the target state is assumed to be equal to $\zeta_0$, \emph{i.e.} to the squeezing of the first spectral supermode ($k = 0$) and set to give -3 dB of noise reduction with respect to the standard quantum level; such a value is commonly used in Schr\"odinger kitten generation~\cite{LvovskyNG2020}. Fig.~\ref{CasoReale}-\emph{bottom} shows, for different $K$, the LO bandwidth and the filter FWHM required to obtain a fidelity with the target state of 0.95. As it can be seen for $K$=9, this leads to a filter FWHM of $\approx$\,1.15\,nm and to a $\Delta \lambda_{LO}\approx$\,2.15\,nm.


 \section*{Conclusions}
This work presents a versatile and practical theoretical framework allowing to describe single-photon subtraction in the case where no mode-selective operation is available and a simple spectral filter is considered on the heralding path of the scheme. Such configuration considerably simplifies the experimental setups, making them easier to be implemented. The explicit shape of the heralded state, as measured by a homodyne detector, is provided as a function of the multimode character of the protocol input state, of the spectral profile of the filter and of the local oscillator shape. Based on the desired target features, this allows choosing adequate working conditions for future practical realisations, thus representing a powerful tool for optimally designing future experiments. The strength of the model is illustrated by applying it to the generation of Schr\"odinger kittens, obtained from single-photon subtraction from a multimode squeezed state. Different working conditions have been investigated and discussed, to determine the features of heralded non-Gaussian states that can be actually obtained and retrieved when single-mode generation or state manipulation are not easy to be implemented. 
	
 \section*{ACKNOWLEDGMENTS}

This work has been conducted within the framework of the project HyLight (No. ANR-17-CE30-0006-01) granted by the Agence Nationale de la Recherche (ANR).

\appendix
\section{Properties of functions \texorpdfstring{$\{\psi_{n}^{/\!\!/}(\omega),\psi_{n}^{\perp}(\omega)\}$}{}  }
\label{mammt}
In this appendix, we provide some important properties of the functions $\{\psi_{n}^{/\!\!/}(\omega)\}$ and $\{\psi_{n}^{\perp}(\omega)\}$ that are used in the main text. The starting point consists in decomposing the functions $t(\omega)\psi_{k}(\omega)$ in an orthonormal basis $\{\psi_{n}^{/\!\!/}(\omega)\}$:
\begin{align}
t(\omega)\psi_{k}(\omega)=\sum_n a_{kn}\psi_{n}^{/\!\!/}(\omega),
\label{othonormalization}
\end{align}
with
\begin{align}
a_{kn}=\int t(\omega)\psi_{k}(\omega)\left(\psi_{n}^{/\!\!/}(\omega)\right)^*d\omega,
\end{align}
and to complete  this basis with the set $\{\psi_{n}^{\perp}(\omega)\}$. The functions  $\{\psi_{n}^{\perp}(\omega)\}$ are orthogonal to each other and to the $\{\psi_{n}^{/\!\!/}(\omega)\}$, such that the ensemble of $\{\psi_{n}^{/\!\!/}(\omega), \psi_{n}^{\perp}(\omega)\}$ forms a complete orthonormal basis.
By inverting the relations~\eqref{pandp}, one obtains:
\begin{subequations}
\begin{align}
\psi_{n}^{/\!\!/}(\omega)&=\sum_kp_{kn}^*\psi_k(\omega)\label{parmodes_funcinput}\\
\psi_{n}^{\perp}(\omega)&=\sum_kq_{kn}^*\psi_k(\omega)\label{orthmodes_funcinput}.
\end{align}
\end{subequations}

Also, by noting that
\begin{subequations}
\begin{align}
\sum_kp_{kn_1}p_{kn_2}^*=\sum_kq_{kn_1}q_{kn_2}^*=\delta_{n_1,n_2},\\
\sum_kp_{kn_1}q_{kn_2}^*=\sum_kq_{kn_1}p_{kn_2}^*=0,
\end{align}
\end{subequations}
it can be proved, thanks to (\ref{eq_basis_change}) and 
\begin{align}
\sum_k\psi_k(\omega)\psi_k^*(\omega')=\delta(\omega-\omega')\end{align}
that
\begin{align}
\quad\sum_l\psi_l^{/\!\!/}(\omega)\left(\psi_l^{/\!\!/}(\omega')\right)^*+\psi_l^{\perp}(\omega)\left(\psi_l^{\perp}(\omega')\right)^*=\delta(\omega-\omega'),
\label{Completeness_new}
\end{align}
thus providing a completeness relations for the ensemble of functions $\{\psi_{n}^{/\!\!/}(\omega),\psi_{n}^{\perp}(\omega)\}$.

By multiplying (\ref{parmodes_funcinput}) and (\ref{orthmodes_funcinput}) by $t(\omega)$ and remembering definition (\ref{othonormalization}) one obtains:
\begin{subequations}
\begin{align}
t(\omega)\psi_{n}^{/\!\!/}(\omega)&=\sum_l\left(\sum_kp_{kn}^*a_{kl}\right)\psi_{l}^{/\!\!/}(\omega)=\sum_l T_{nl}\psi_{l}^{/\!\!/},\\
t(\omega)\psi_{n}^{\perp}(\omega)&=\sum_l\left(\sum_kq_{kn}^*a_{kl}\right)\psi_{l}^{/\!\!/}(\omega).
\end{align}
\label{trans_transformodes}
\end{subequations}
Then:
\begin{align}
\int|t(\omega)|^2|\psi_n^{\perp}(\omega)|^2d\omega=\int\left[|t(\omega)|^2\psi_n^{\perp}(\omega)\right]\left(\psi_n^{\perp}(\omega)\right)^*d\omega.
\end{align}
However,  according to the two relations (\ref{trans_transformodes}), $|t(\omega)|^2\psi_n^{\perp}(\omega)$ only has components along $\psi_{n}^{/\!\!/}$, which implies that $\int|t(\omega)|^2|\psi_n^{\perp}(\omega)|^2d\omega=0$, and, as a consequence, that:
\begin{align}
\forall \omega,\quad t(\omega)\psi_n^{\perp}(\omega)=0.\label{psiperp_tran}
\end{align}
This result is quite strong, and implies that \textbf{$\psi_n^{\perp}(\omega)$ is only nonzero where $r(\omega)=1$}, \emph{i.e.} in all region for which the filter transmission coefficient $t(\omega)$ goes exactly to 0. This also means that 
\begin{align}
\forall\omega,\ r(\omega)\ \psi_n^{\perp}(\omega)=\psi_n^{\perp}(\omega).
\label{psiperp_refl}
\end{align}
Note also that (\ref{othonormalization}) implies that if $t(\omega)=0$ then $\psi_{n}^{/\!\!/}(\omega)=0$, for regular enough continuous functions, with continuous derivatives, such as for a Gaussian or Lorentzian filter transmission profile. This implies that $\psi_n^{\perp}(\omega)$ and $\psi_{n}^{/\!\!/}(\omega)$ have disjoint supports.
\section{Fidelity of the measured state with a target non-Gaussian state}\label{appendice2}
The quality of the detected state can be evaluated by its fidelity with a monomode photon subtracted squeezed vacuum state as target state thaks to the overlap formula:
\begin{equation}\label{Fidelity}
\mathcal{F} = 2\pi \int\int W_{H}(x, y)W_{t}(x, y) dx dy.
\end{equation}

By replacing \eqref{Multimode_filtered_Wigner_function} and \eqref{targetammammeta} in (\ref{Fidelity}), we find that:
\begin{align}
\mathcal{F} =& \dfrac{2}{P \sigma_x \sigma_y \sqrt{(\frac{1}{\sigma_x^2} + \frac{2}{s})(\frac{1}{\sigma_y^2} + 2s)}} \left[ -A +\right. \nonumber\\
& \left( \dfrac{2A}{s} - B \right) \dfrac{1}{\frac{1}{\sigma_x^2} + \frac{2}{s}} +\left( 2As - C \right) \dfrac{1}{\frac{1}{\sigma_y^2} + 2s} 
 \nonumber\\
 & +  \dfrac{2B}{s} \dfrac{3}{(\frac{1}{\sigma_x^2} + \frac{2}{s})^2}+ 2Cs \dfrac{3}{(\frac{1}{\sigma_y^2} + 2s)^2} \nonumber\\
 &\left.\left( \dfrac{2C}{s} + 2Bs \right)\dfrac{1}{(\frac{1}{\sigma_x^2} + \frac{2}{s})(\frac{1}{\sigma_y^2} + 2s)} \right],
\end{align}

with
\begin{subequations}
\begin{align}
A =& P - \dfrac{1}{2\sigma_x^2}\sum_{k, n} \gamma_{k,n}\dfrac{\mu_k\mu_n c_k c_n}{(1 - \mu_k)(1 - \mu_n)} -\nonumber\\
& \dfrac{1}{2\sigma_y^2}\sum_{k, n} \gamma_{k,n}\dfrac{\mu_k\mu_n c_k c_n}{(1 + \mu_k)(1 + \mu_n)}
\end{align}

\begin{align}
B = \dfrac{1}{2\sigma_x^4}\sum_{k, n} \gamma_{k,n}\dfrac{\mu_k\mu_n c_k c_n}{(1 - \mu_k)(1 - \mu_n)}
\end{align}

\begin{align}
C = \dfrac{1}{2\sigma_y^4}\sum_{k, n} \gamma_{k,n}\dfrac{\mu_k\mu_n c_k c_n}{(1 + \mu_k)(1 + \mu_n)}.
\end{align}
\end{subequations}

\section{Gaussian and rectangular filters on the heralding path}\label{Appendice3}
In the case of a Gaussian shaped JSA, the explicit shape of the supermode is given as in~\cite{wasilewski2006pulsed}:
\begin{equation}\label{Signal_modes}
\psi_k (\omega) \propto \sqrt{\dfrac{\tau_s}{\sqrt{\pi}2^k k!}} H_k\left[ \tau_s (\omega - \omega_{p}/2) \right] e^{-\tau_s^2 (\omega - \omega_{p}/2)^2/2},
\end{equation}
where $\omega_p$ is the SPDC pump central frequency and $\tau_s$ is given by the pump and by the process $K$~\cite{wasilewski2006pulsed}. In general, 
\begin{equation}
\gamma_{k,n} = \int \vert t(\omega) \vert^2 \psi_{k}(\omega) \psi_{n}^*(\omega) d\omega,
\end{equation}
where $\psi_{k}(\omega)$ is the signal mode $k$ given by the equation \ref{Signal_modes} with a phase factor which does not depend on the mode order $k$ \cite{wasilewski2006pulsed}. 

A Gaussian filter with transmittance $t(\omega)$ can be described by:
\begin{equation}
t(\omega) = e^{-\frac{4 \ln (2)}{\Delta\omega_F^2}(\omega - \frac{\omega_p}{2})^2},
\end{equation}
where $\Delta\omega_F$ is the FWHM bandwidth. Accordingly, by defining $\bar{\tau}^2 = \dfrac{8 \ln (2)}{\Delta\omega_F^2} + \tau_s^2$, the coefficients $\gamma_{k,n}$ are:
\begin{align}
&\gamma_{k,n} = \dfrac{\tau_s}{\sqrt{2^{k + n} k! n! \pi}}\times\nonumber\\
& \int e^{-\bar{\tau}^2(\omega -\frac{\omega_p}{2})^2} H_k\left[ \tau_s (\omega - \frac{\omega_p}{2}) \right] H_m\left[ \tau_s (\omega - \frac{\omega_p}{2}) \right] d\omega.
\end{align}
By using the integration formula given in \cite{gradshteyn2014table}:
\begin{align}
&\int e^{-y^2} H_k(a y) H_n(a y) dy =\nonumber\\
& \sqrt{\pi} \sum_{m = 0}^{min[k, n]} 2^m m! \binom{k}{m} \binom{n}{m} (1 - a^2)^{\frac{k + n}{2} - m} H_{k + n - 2m}(0),
\end{align}
it is possible to write:
\begin{align}
\gamma_{k,n} 
&= \dfrac{\tau_s}{\sqrt{\bar{\tau}^2 2^{k + n} k! n!}}\times\nonumber\\
& \sum_{m = 0}^{min[k, n]} 2^m m! \binom{k}{m} \binom{n}{m} (1 - \dfrac{\tau_s^2}{\bar{\tau}^2})^{\frac{k + n}{2} - m} H_{k + n - 2m}(0).
\end{align}
As the Hermite-Gauss polynomials verify: 
\begin{equation}
H_k (0) = \left\lbrace \begin{array}{ll}
0  &\text{if k is odd} \\
(-1)^{\frac{k}{2}} \dfrac{k!}{(\frac{k}{2})!}  &\text{if k is even.} \end{array}\right.
\end{equation}
Then, for a Gaussian filter, $\gamma_{k,n} = 0$ if $k + n$ is odd, and 
\begin{align}
&\gamma_{k,n} =\dfrac{\tau_s}{\sqrt{\nu 2^{k + n} k! n!}}\times\nonumber\\
& \sum_{m = 0}^{min[k, n]} 2^m m! \binom{k}{m} \binom{n}{m} (\dfrac{\tau_s^2}{\nu} - 1)^{\frac{k + n}{2} - m} \dfrac{(k + n - 2m)!}{(\frac{k + n}{2} - m)!}
\end{align}
if $k + n$ is even.\\
Following a similar reasoning, the case of a rectangular filter can also be described. In this case:
\begin{equation}
t(\omega) = \left\lbrace \begin{array}{lll}
T_0  & \text{if} &  \omega \in [-\frac{\Delta \omega_F}{2}, +\frac{\Delta \omega_F}{2}] \\
0  & &\text{elsewhere.}
\end{array}\right.
\end{equation}
In this case, $\gamma_{k,n}=0$ if $k+n$ is odd, and:
\begin{align}
\gamma_{k,n} =
\dfrac{1}{\sqrt{2^{k + n} k! n!}} \int_{-\tau_s \frac{\Delta \omega_F}{2}}^{+\tau_s \frac{\Delta \omega_F}{2}}e^{-x^2}H_k(x)H_n(x) dx
\end{align}
if $k+n$ is even.
\section{Heralded single photon production in a non mode-selective configuration}\label{Appendice4}
The impact of the filter bandwidth on heralded state preparation can be discussed in the context of heralded single-photon sources. In this case, a particularly common configuration relies on pairs of single photons produced by a SPDC process in the weak pumping regime~\cite{branczyk2010optimized}; by deterministically separating the paired photons, for instance thanks to their polarisation, it is indeed possible to use the detection of one photon to herald the presence of its twin. The quality of produced states strongly depends on the possibility of engineering the SPDC process so as to work in single-mode configuration ($K$=1), a condition that often corresponds to sources exploiting type II SPDC~\cite{branczyk2010optimized}. Here, we will consider instead the case of degenerate SPDC, such that paired photons have parallel polarisations and that two-mode SPDC is not experimentally accessible; this configuration is for instance very common in integrated quantum optics~\cite{Mondain2019, Lenzini}. In the special case of weak pumping approximation, light at the output of the degenerate SPDC can be approximated as in the state: 
\begin{equation}
\ket{\psi}\approx \ket{0}+A\iint d\omega d\omega' f(\omega, \omega') \hat{a}^\dagger(\omega)\hat{a}^\dagger(\omega'),
\end{equation}
where the joint spectral amplitude $f(\omega, \omega')$ (JSA) depends on the SPDC pump spectral width and on the process working condition via the phase-matching. The factor $A$ gathers the constants linked with the non linear process, such as pump power and non linear susceptibility (here $A<<1$ as we are in the weak pumping regime). The bosonic operators verify the commutation rule $[\hat{a}(\omega), \hat{a}^\dagger(\omega')]=\delta (\omega-\omega')$ and $[\hat{a}(\omega), \hat{a}(\omega')]=0$. The special case of perfectly single-mode regime (\emph{i.e.} $K$=1) corresponds to a factorisable JSA, $f(\omega, \omega')=g(\omega)h(\omega')$~\cite{wasilewski2006pulsed}. The state after the subtraction beamsplitter and the filtering stage can be obtained by employing the expressions~\eqref{BS_sub} and~\eqref{Beamsplitt}. In the special case considered here, dealing with single photon regime, vacuum contributions are traced out by the single-photon detection process and the state right before the detectors can be written as:
\begin{widetext}
\begin{equation}
\ket{\psi'}\propto \iiint d\omega d\omega'd\omega'' f(\omega, \omega') t(\omega'') \hat{a}(\omega'')\hat{b}^{\dagger}_{\rm out}(\omega'')\hat{a}^\dagger(\omega)\hat{a}^\dagger(\omega')\ket{0},
\end{equation}
\end{widetext}
where $\hat{b}_{\rm out}(\omega)$ indicates spectral components at frequency $\omega$ downstream of the filter defined as in Eq.\eqref{boutcout}. As in the main text, the heralded state can be obtained by introducing a POVM operator describing the detection of light at the SPD. In the case considered here, without introducing the description in terms of the supermodes, the single photon detection is associated with the POVM operator:
\begin{equation}
\hat{\Pi}_{\rm out}=\int d\omega \ketbra{1_{\rm out}(\omega)},
\end{equation}
where $\ket{1_{\rm out}(\omega)}=\hat{b}^{\dagger}_{\rm out}(\omega)\ket{0}$ and describes, in a very general way, single photons downstream of the filter. By tracing over the degrees of freedom of the detected mode (\emph{i.e.} those related to $\hat{b}_{\rm out}$), the heralded state can be written as:
\begin{widetext}
\begin{equation}
\hat{\rho}_{\rm out}\propto\int d\omega d\omega' d\omega''d\Omega d\Omega' f(\omega, \omega')t(\omega'')\hat{a}(\omega'')\hat{a}^\dagger(\omega)\hat{a}^\dagger(\omega')\ketbra{0}
\hat{a}(\Omega)\hat{a}(\Omega')\hat{a}^\dagger(\omega'')t^*(\omega'')f^*(\Omega,\Omega').
\end{equation} 
\end{widetext}
A particularly interesting case is represented by the limit of an extremely narrowband filter, that we describe here by a transmission profile $t(\omega)=\delta (\omega-\omega_0)$. In this limit, the previous expression reads as:
\begin{widetext}
\begin{equation}
\hat{\rho}_{\rm out}\propto\int d\omega d\omega'd\Omega d\Omega' f(\omega, \omega')\hat{a}(\omega_0)\hat{a}^\dagger(\omega)\hat{a}^\dagger(\omega')\ketbra{0}
\hat{a}(\Omega)\hat{a}(\Omega')\hat{a}^\dagger(\omega_0)f^*(\Omega,\Omega').
\end{equation} 
\end{widetext}
By exploiting the commutation relations among bosonic operators at different frequency, it is possible to re-express the output state as: 
\begin{widetext}
\begin{align}
\hat{\rho}_{\rm out}\propto&\int d\omega d\omega' [f(\omega, \omega_0)f^*(\omega_0, \omega')+f(\omega_0, \omega)f^*(\omega_0, \omega')]\hat{a}^\dagger(\omega)\ketbra{0}\hat{a}^\dagger(\omega')\nonumber\\
&+\int d\omega d\omega' [f(\omega, \omega_0)f^*(\omega', \omega_0)+f(\omega_0, \omega)f^*(\omega', \omega_0)]\hat{a}^\dagger(\omega)\ketbra{0}\hat{a}^\dagger(\omega')\nonumber\\
=&\int d\omega d\omega' G(\omega)G^*(\omega')\hat{a}^\dagger(\omega)\ketbra{0}\hat{a}^\dagger(\omega'),
\label{rhofinalediagosto}
\end{align}
\end{widetext}
where we introduced $G(\omega)=f(\omega,\omega_0)+f(\omega_0,\omega)$. By defining 
$\hat{A}=\int d\omega G(\omega) \hat{a}^\dagger (\omega)$, Eq.~\eqref{rhofinalediagosto} can be re-expressed as: 
\begin{equation}
\hat{\rho}_{\rm out}\propto \hat{A}^\dagger \ketbra{0}\hat{A}.
\end{equation} 
This relation correctly describes a pure single photon state heralded in the mode described by the operator $\hat{A}$. This confirms the results found in the very general case described in the main text, about the purity and the quality of heralded state with narrowband filters. This result provides an analytical proof that optimal filtering is as narrow as possible for the generation of odd Schr\"odinger cat states with vanishingly small amplitude (single photon Fock states). This reasoning must be put in parallel with the well-established result that a pure signal photon can be heralded by the detection of a narrow filtered idler photon from a non-degenerate SPDC source.

\bibliography{biblioMultiCats}

\begin{thebibliography}{10}

\bibitem{Giovannetti2006}
V.~Giovannetti, S.~Lloyd, and L.~Maccone, ``Quantum metrology,'' {\em Phys.
  Rev. Lett.}, vol.~96, p.~010401, 2006.

\bibitem{Braunstein2005a}
S.~L. Braunstein and P.~van Loock, ``Quantum information with continuous
  variables,'' {\em Rev. Mod. Phys.}, vol.~77, p.~513, 2005.

\bibitem{Ferraro_Book}
A.~Ferraro, S.~Olivares, and M.~G.~A. Paris, {\em Gaussian states in continuous
  variable quantum information}.
\newblock Bibliopolis, Napoli, 2005.

\bibitem{ShapiroGaussianQI2012}
C.~Weedbrook, S.~Pirandola, R.~Garc\'{\i}a-Patr\'on, N.~J. Cerf, T.~C. Ralph,
  J.~H. Shapiro, and S.~Lloyd, ``Gaussian quantum information,'' {\em Rev. Mod.
  Phys.}, vol.~84, pp.~621--669, 2012.

\bibitem{dakna1997generating}
M.~Dakna, T.~Anhut, T.~Opatrn{\`y}, L.~Kn{\"o}ll, and D.-G. Welsch,
  ``Generating schr{\"o}dinger-cat-like states by means of conditional
  measurements on a beam splitter,'' {\em Physical Review A}, vol.~55, no.~4,
  p.~3184, 1997.

\bibitem{LvovskyNG2020}
A.~I. Lvovsky, P.~Grangier, A.~Ourjoumtsev, V.~Parigi, M.~Sasaki, and
  R.~Tualle-Brouri, ``{Production and applications of non-Gaussian quantum
  states of light},'' {\em arXiv:2006.16985v1}, 2020.

\bibitem{ourjoumtsev2006}
A.~Ourjoumtsev, R.~Tualle-Brouri, J.~Laurat, and P.~Grangier, ``Generating
  optical schr{\"o}dinger kittens for quantum information processing,'' {\em
  Science}, vol.~312, no.~5770, pp.~83--86, 2006.

\bibitem{neergaard2006}
J.~S. Neergaard-Nielsen, B.~M. Nielsen, C.~Hettich, K.~M{\o}lmer, and E.~S.
  Polzik, ``Generation of a superposition of odd photon number states for
  quantum information networks,'' {\em Physical review letters}, vol.~97,
  no.~8, p.~083604, 2006.

\bibitem{wakui2007}
K.~Wakui, H.~Takahashi, A.~Furusawa, and M.~Sasaki, ``Photon subtracted
  squeezed states generated with periodically poled ktiopo 4,'' {\em Optics
  Express}, vol.~15, no.~6, pp.~3568--3574, 2007.

\bibitem{wasilewski2006pulsed}
W.~Wasilewski, A.~Lvovsky, K.~Banaszek, and C.~Radzewicz, ``Pulsed squeezed
  light: Simultaneous squeezing of multiple modes,'' {\em Physical Review A},
  vol.~73, no.~6, p.~063819, 2006.

\bibitem{sasaki2006multimode}
M.~Sasaki and S.~Suzuki, ``{Multimode theory of measurement-induced
  non-Gaussian operation on wideband squeezed light: Analytical formula},''
  {\em Physical Review A}, vol.~73, no.~4, p.~043807, 2006.

\bibitem{Silberhorn2011SingleModeTWB}
A.~Eckstein, A.~Christ, P.~J. Mosley, and C.~Silberhorn, ``{Highly Efficient
  Single-Pass Source of Pulsed Single-Mode Twin Beams of Light},'' {\em Phys.
  Rev. Lett.}, vol.~106, p.~013603, 2011.

\bibitem{walschaers2018tailoring}
M.~Walschaers, S.~Sarkar, V.~Parigi, and N.~Treps, ``Tailoring non-gaussian
  continuous-variable graph states,'' {\em Physical review letters}, vol.~121,
  no.~22, p.~220501, 2018.

\bibitem{TrepsNG2020}
Y.-S. Ra, A.~Dufour, M.~Walschaers, C.~Jacquard, T.~Michel, C.~Fabre, and
  N.~Treps, ``{Non-Gaussian quantum states of a multimode light field},'' {\em
  Nat. Phys.}, vol.~16, pp.~144--147, 2020.

\bibitem{averchenko2016multimode}
V.~Averchenko, C.~Jacquard, V.~Thiel, C.~Fabre, and N.~Treps, ``Multimode
  theory of single-photon subtraction,'' {\em New Journal of Physics}, vol.~18,
  no.~8, p.~083042, 2016.

\bibitem{QuantumPulseGate}
A.~Eckstein, B.~Brecht, and C.~Silberhorn, ``A quantum pulse gate based on
  spectrally engineered sum frequency generation,'' {\em Opt. Express},
  vol.~19, pp.~13770--13778, Jul 2011.

\bibitem{Mondain2019}
F.~Mondain, T.~Lunghi, A.~Zavatta, E.~Gouzien, F.~Doutre, M.~{De Micheli},
  S.~Tanzilli, and V.~D'Auria, ``{Chip-based squeezing at a telecom
  wavelength},'' {\em Photonics Research}, vol.~7, no.~7, pp.~A36--A39, 2019.

\bibitem{Patera2010}
G.~Patera, N.~Treps, C.~Fabre, and G.~J. de~Valc\'{a}rcel, ``{Quantum theory of
  synchronously pumped type I optical parametric oscillators: generation of
  multiple, squeezed frequency combs below threshold},'' {\em Eur. Phys. J. D},
  vol.~56, p.~123, 2010.

\bibitem{Patera2020}
E.~Gouzien, S.~Tanzilli, V.~D'Auria, and G.~Patera, ``{Morphing Supermodes: A
  Full Characterization for Enabling Multimode Quantum Optics},'' {\em Phys.
  Rev. Lett.}, vol.~125, p.~103601, 2020.

\bibitem{branczyk2010optimized}
A.~M. Bra{\'n}czyk, T.~Ralph, W.~Helwig, and C.~Silberhorn, ``{Optimized
  generation of heralded Fock states using parametric down-conversion},'' {\em
  New Journal of Physics}, vol.~12, no.~6, p.~063001, 2010.

\bibitem{christ2014theory}
A.~Christ, C.~Lupo, M.~Reichelt, T.~Meier, and C.~Silberhorn, ``{Theory of
  filtered type-II PDC in the continuous-variable domain: Quantifying the
  impacts of filtering},'' {\em Phys. Rev. A}, vol.~90, p.~023823, 2014.

\bibitem{Furusawa2017CWCats}
W.~Asavanant, K.~Nakashima, Y.~Shiozawa, J.-I. Yoshikawa, and A.~Furusawa,
  ``{Generation of highly pure Schroedinger's cat states and real-time
  quadrature measurements via optical filtering},'' {\em Opt. Express},
  vol.~25, pp.~32227--32242, 2017.

\bibitem{GattiOriginali2007}
A.~Ourjoumtsev, ``{Theoretical and experimental study of quantum coherent
  superpositions and of non-Gaussian entangled states of the light},'' {\em PhD
  Thesis}, 2008.

\bibitem{nielsen2010quantum}
M.~A. Nielsen and I.~L. Chuang, {\em Quantum Computation and Quantum
  Information}.
\newblock Cambridge University Press, 2010.

\bibitem{Gouz2018Time}
E.~Gouzien, B.~Fedrici, A.~Zavatta, S.~Tanzilli, and V.~D'Auria, ``Quantum
  description of timing jitter for single-photon on-off detectors,'' {\em Phys.
  Rev. A}, vol.~98, p.~013833, Jul 2018.

\bibitem{Laurat2012EPJD}
V.~D'Auria, O.~Morin, C.~Fabre, and J.~Laurat, ``Effect of the heralding
  detector properties on the conditional generation of single-photon states.,''
  {\em Eur. Phys. J. D}, vol.~66, p.~249, 2012.

\bibitem{Christ2011}
A.~Christ, B.~Brecht, W.~Mauerer, and C.~Silberhorn, ``Probing multimode
  squeezing with correlation functions,'' {\em New J. Phys.}, vol.~13,
  p.~033027, 2011.

\bibitem{hudson1974wigner}
R.~L. Hudson, ``When is the wigner quasi-probability density non-negative?,''
  {\em Reports on Mathematical Physics}, vol.~6, no.~2, pp.~249--252, 1974.

\bibitem{genoni2013detecting}
M.~G. Genoni, M.~L. Palma, T.~Tufarelli, S.~Olivares, M.~Kim, and M.~G. Paris,
  ``Detecting quantum non-gaussianity via the wigner function,'' {\em Physical
  Review A}, vol.~87, no.~6, p.~062104, 2013.

\bibitem{filip2011detecting}
R.~Filip and L.~Mi{\v{s}}ta~Jr, ``Detecting quantum states with a positive
  wigner function beyond mixtures of gaussian states,'' {\em Physical review
  letters}, vol.~106, no.~20, p.~200401, 2011.

\bibitem{kenfack_2004_negativity}
A.~Kenfack and K.~{\.Z}yczkowski, ``{Negativity of the Wigner function as an
  indicator of non-classicality},'' {\em Journal of Optics B: Quantum and
  Semiclassical Optics}, vol.~6, no.~10, p.~396, 2004.

\bibitem{Lenzini}
F.~Lenzini, J.~Janousek, O.~Thearle, M.~Villa, B.~Haylock, S.~Kasture, L.~Cui,
  H.-P. Phan, D.~V. Dao, H.~Yonezawa, P.~K. Lam, E.~H. Huntington, and
  M.~Lobino, ``Integrated photonic platform for quantum information with
  continuous variables,'' {\em Science Advances}, vol.~4, no.~12, 2018.

\bibitem{Matthews2021}
J.~F. Tasker, J.~Frazer, G.~Ferranti, E.~J. Allen, L.~F. Brunel, S.~Tanzilli,
  V.~D’Auria, and J.~C. Matthews, ``Silicon photonics interfaced with
  integrated electronics for 9 ghz measurement of squeezed light,'' {\em Nature
  Photonics}, vol.~15, no.~1, pp.~11--15, 2021.

\bibitem{gradshteyn2014table}
I.~S. Gradshteyn and I.~M. Ryzhik, {\em Table of integrals, series, and
  products}.
\newblock Academic press, 2014.

\end{thebibliography}
\bibliographystyle{ieeetr}

\end{document}